\documentclass[aps,pra,showpacs,notitlepage,onecolumn,superscriptaddress]{revtex4-1}
\usepackage[utf8]{inputenc}
\usepackage[tmargin=1in, bmargin=1in, lmargin=1.25in, rmargin=1.25in]{geometry}
\usepackage{amsmath, amssymb, amsthm}
\usepackage{graphicx}
\usepackage{xcolor}
\usepackage{enumitem}
\usepackage{titlesec}
\usepackage{datetime}
\usepackage{hyperref}
\usepackage{import}
\usepackage{mathtools}

\usepackage{tikz-cd}

\usepackage{kbordermatrix}

\definecolor{crimson}{RGB}{186,0,44}

\hypersetup{
    colorlinks,
    linkcolor={crimson},
    citecolor={crimson},
    urlcolor={crimson}
}

\theoremstyle{definition}
\newtheorem{definition}{Definition}[section]
\newtheorem{lemma}{Lemma}[section]
\newtheorem{theorem}{Theorem}[section]
\newtheorem{corollary}{Corollary}[theorem]
\newtheorem*{theorem*}{Theorem}
\newtheorem*{corollary*}{Corollary}
\newtheorem{remark}{Remark}[section]

\newtheorem{example}{Example}[section]

\newtheorem{problem}{Problem}[section]

\begin{document}

\title{Semantic embedding for quantum algorithms}
\author{Zane M.\ Rossi}
\email{zmr@mit.edu}
\affiliation{%
Department of Physics,
Massachusetts Institute of Technology, Cambridge, Massachusetts 02139, USA}
\author{Isaac L.\ Chuang}\affiliation{%
Department of Physics,
Massachusetts Institute of Technology, Cambridge, Massachusetts 02139, USA}
\date{\today}

\begin{abstract}
    \noindent The study of classical algorithms is supported by an immense understructure, founded in logic, type, and category theory, that allows an algorithmist to reason about the sequential manipulation of data irrespective of a computation's realizing dynamics. 
    As quantum computing matures, a similar need has developed for an assurance of the correctness of high-level quantum algorithmic reasoning.
    Parallel to this need, many quantum algorithms have been unified and improved using quantum signal processing (QSP) and quantum singular value transformation (QSVT), which characterize the ability, by alternating circuit ansätze, to transform the singular values of sub-blocks of unitary matrices by polynomial functions.
    However, while the algebraic manipulation of polynomials is simple (e.g., compositions and products), the QSP/QSVT circuits realizing analogous manipulations of their embedded polynomials are non-obvious.
    This work constructs and characterizes the runtime and expressivity of QSP/QSVT protocols where circuit manipulation maps naturally to the algebraic manipulation of functional transforms (termed semantic embedding).
    In this way, QSP/QSVT can be treated and combined modularly, purely \emph{in terms of the functional transforms they embed}, with key guarantees on the computability and modularity of the realizing circuits.
    We also identify existing quantum algorithms whose use of semantic embedding is implicit, spanning from distributed search to proofs of soundness in quantum cryptography. The methods used, based in category theory, establish a theory of semantically embeddable quantum algorithms, and provide a new role for QSP/QSVT in reducing sophisticated algorithmic problems to simpler algebraic ones.
\end{abstract}

\maketitle

\vspace{-2em}



\section{Introduction}

\noindent It is an understated but core tenet of computing that the semantic structure of an algorithm can be made independent of its representation. That is, while an algorithmist blithely concerns themselves with the sequential application of functions to meaningful data, said algorithm is in truth run on a computational system whose structure and initialization might bear little resemblance to the considered problem. That algorithms can be reasoned about so agnostically to the systems in which they are embedded is the product of enormous effort in the development of classical computer languages and methods for verifying program meaning and correctness \cite{gunter_semantics_92, hoare_axiomatic_69, floyd_program_meaning_93, apt_verification_09}. Analogously, as systems for quantum information processing become increasingly sophisticated, the need has grown for similar representation-agnosticism in quantum computing \cite{selinger_qpl_04, gay_qpls_06, ac_category_semantics_04}.

Building on methods from classical computation, this work takes steps toward specifying how quantum algorithms with semantic structure can, as subroutines, be put together while ensuring the preservation of this structure. Specifically, we focus on the combination of instances of quantum signal processing (QSP) and quantum singular value transformation (QSVT), which have recently had success in unifying, simplifying, and improving most known quantum algorithms \cite{mrtc_21, gslw_19}. These algorithms perform polynomial transformations on the singular values of embedded linear operators, and consequently the semantic structure we endeavor to preserve in their combination relates strongly to these polynomial transforms. Beyond its value in the design of complex quantum algorithms, we are also able to show that combining QSVT subroutines in this way has application to distributed scheduling protocols \cite{grover_05}, distributed linear systems problems \cite{ms_communication_complexity_22}, and proofs of security against quantum adversaries in cryptographic protocols \cite{cmsz_qpsa_2022, lombardi_pqzk_2021}.

Unsurprisingly, preserving the semantic structure of QSVT circuits as subroutines under combination imposes strict requirements on the form of this combination. It is known that a variety of ubiquitous concepts in classical control flow \cite{ying_control_flow_12} and recursion \cite{xyv_recursive_programs_21} cannot be naïvely applied to quantum computations, and thus these results provide methods for clear, useful, and above all simply achievable manipulations at the algorithmic level for quantum information processing. To the extent that QSVT can achieve state-of-the-art performance for most known quantum algorithmic tasks, and under known restrictions on the functional transformations they embed, this work fully characterizes the rules under which such modules can be (algorithmically) meaningfully combined. The understructure of this theory is built from, as per its classical counterpart \cite{moggi_cat_91, tp_semantics_97}, basic but elegant techniques in category theory \cite{cat_theory_78}, suggesting an exciting prospect for the expanded applicability of this subfield to the theory of quantum computation.

While this paper is not solely for those familiar with QSP and QSVT, the specificity of its problem statement and methods requires some knowledge of how these quantum algorithms work. As such, in the following paragraphs we discuss high-level properties of QSP and QSVT, toward introducing a category-theoretic diagram, Fig.~\ref{fig:qsp_as_natural_transformation}, summarizing our core contribution. We aim to keep this discussion simple, generic, and intuition-forward. As such, we want to dissuade a view of QSP and QSVT as monolithic or fixed algorithms, promoting them instead as concrete instances of a wider, \emph{modular} approach to quantum information processing.

Quantum signal processing (QSP) \cite{ylc_14, lyc_16_equiangular_gates, lc_17_simultation, lc_19_qubitization} and quantum singular value transformation (QSVT) \cite{gslw_19}, quantum algorithms for obliviously transforming the singular values of linear operators encoded in unitary matrices by polynomial functions, have enjoyed recent success in unifying and improving most quantum algorithms \cite{gslw_19, mrtc_21, cs_qsvt_tang_tian}. This unification is founded in the surprising expressivity of the QSVT circuit ansatz, which permits one to cleanly take algorithmic problems to questions about the existence of polynomial functions with desired properties. Despite this success, the applicability of QSP and QSVT in generating new quantum algorithms remains limited by the opaqueness of the map between quantum circuits and their embedded polynomial transformation. 

The action of QSP is to take access to an oracle $W(x) \in \text{SU}(2)$ parameterized by a scalar value $x \in \mathbb{R}$ (the signal) and output a unitary circuit that embeds a polynomial transformation $P(x) \in \mathbb{C}[x]$. Crucially, by `embeds' we mean (with concrete definitions to follow) that this $P(x)$ relates simply to the distribution assigned to a simple observable, allowing measurements to depend strongly on properties of the signal.

The abstraction of a QSP circuit to the polynomial it embeds is so useful in fact that it is tempting to think of QSP/QSVT protocols \emph{solely in terms of the functional transforms they embed}. These transforms are imbued with, and interpretable according to, algorithmic tasks (they are \emph{semantic}). E.g., for amplitude amplification, one desires a polynomial which is approximately constant over its domain \cite{mrtc_21}. Moreover, the cost of the implementation of these polynomials, their action under noise, and the stability of their numerical computation, are well-understood, and grounded in long-standing functional analytic techniques. The business of this paper is to modify this temptation to view QSP/QSVT `function-first' into a firm, mathematical statement.

A natural question is whether a larger class of quantum algorithmic tasks than QSP/QSVT can be viewed through a \emph{purely functional lens}. Formalizing this question is a matter of understanding how algebraic manipulations in \emph{functional space} (the space with semantic interpretation) correspond to their pre-image in the \emph{programmable space} of parameterized circuit ansätze (a space with physical interpretation). If this question is resolvable for a given algebraic manipulation, and moreover if the required map is efficiently computable, then we can perform useful analysis of quantum algorithms while reasoning purely in functional space. This frees the algorithmist to consider higher-order manipulations of quantum information. To summarize the above, we provide the following remark, which reframes known properties of QSP to fit our problem statement.

\begin{remark}[QSP protocols in programmable space and functional space] \label{remark:functional_programmable_spaces}
    A QSP protocol is an alternating quantum circuit ansatz defined by a list of real numbers $\Phi \in \mathbb{R}^{n + 1}$ (the QSP phases), and using a unitary oracle $W(x) \in \text{SU(2)}$ parameterized by some unknown $x \in \mathbb{R}$. The action of this circuit is to take as input $\Phi$ and prepare a unitary that calls $W(x)$ $n$ times and embeds an at most degree $n$ polynomial $P(x)$ (usually as a matrix element), where $P(x)$ relates simply to an observed probability. In other words, QSP has the type signature
        \begin{equation} \label{eq:qsp_type_signature}
            \text{QSP}: \mathbb{R}^{n+1} \rightarrow \mathbb{C}_n[x],
        \end{equation}
    where $\mathbb{C}_n[x]$ are polynomials in $x$ over $\mathbb{C}$ of degree at most $n$, and where we leave the relevant maps and projections unspecified for brevity. In other words, QSP returns an accessible \emph{function} on an unknown parameter. We say here that $\Phi$ sits in \emph{programmable space} (with physical interpretation), while $P$ sits in \emph{functional space} (with semantic interpretation).

    Before introducing a formal problem statement, we can play with the type signature of QSP in Eq.~\ref{eq:qsp_type_signature}; it is easy to imagine maps between sets of real numbers, as well as polynomials. The intent of this work (and a core concept, as explained later, within category theory) is to have such maps remain non-trivial while also meshing with one another. That is, that a graph like the following:
        \begin{equation}
            \begin{tikzcd}
                 \mathbb{R}^{n + 1} \arrow[r, "f"] \arrow[d, "g"] & \mathbb{C}_n[x] \arrow[d, "h"]\\ 
                 \mathbb{R}^{m + 1}  \arrow[r, "f"] & \mathbb{C}_m[x]    
            \end{tikzcd},
        \end{equation}
    permits observations like $h\circ f = f\circ g$, or that one cares only about beginning- and end-points of valid paths according to arrows. In other words, that this diagram \emph{commutes}. The algorithmist here desires mostly to think about $h$, while ensuring that a simple and corresponding $g$ can be found, through an understanding of $f$ (the QSP map from phases to polynomials). $\square$
\end{remark}

Having provided a type signature of the action of QSP (which under slight modification suffices for QSVT, see Eq.~\ref{eq:qsvt_type_signature}), we now can more earnestly use basic ideas in category theory to precisely define what we intend by `analogousness' between manipulations in functional and programmable space. The category-theoretic diagram in Fig.~\ref{fig:qsp_as_natural_transformation} will constitute our problem statement, together with Table~\ref{tab:qsp_categories}, which describes assignments of concrete objects (e.g., circuits and manipulations of circuit parameters) in QSP to the objects and arrows of Fig.~\ref{fig:qsp_as_natural_transformation}.

We intend that the simple form of the problem statement for semantic embedding, given in Fig.~\ref{fig:qsp_as_natural_transformation} through Def.~\ref{def:semantic_embedding}, supports a more ambitious role for QSP and QSVT in understanding quantum algorithms. Further advancements could come in the form of realizing new assignments for functors Fig.~\ref{fig:qsp_as_natural_transformation}, as well as in the analysis of modified QSP-like ansätze. Together, determining how constraints on circuit parameterizations correspond to algorithmic expressivity, and specifying rules for how to semantically combine subroutines while preserving this expressivity, enables and encourages higher-order reasoning about quantum computations in a formal way.

There are substantial barriers in the way of specifying such ansätze and combination rules. As is well known, the map between QSP phases and the achieved polynomial transform is opaque \cite{haah_2019}, and without modification not unique \cite{sym_qsp_21}. Moreover, naïvely recursively nesting a QSP circuit within itself \emph{does not} (as shown after Def.~\ref{def:nested_protocols}) recursively compose the achieved polynomial transform. In other words, \emph{semantic actions on polynomials} do not automatically correspond to \emph{simple manipulations of the QSP phases}. Nevertheless, for simple cases, this correspondence seems to be possible (e.g., Chebyshev polynomials). This work answers positively that a far more expressive set of QSP protocols can be combined semantically as modules. The strength of this work rests in its preservation of the properties that make QSP pleasant: (1) an efficiently computable, numerically stable map from an embedded polynomial to QSP phases, and (2) that a fixed, contiguous QSP protocol can be said to achieve a fixed functional transform. Simply composing polynomials, and only then finding the resulting function's QSP phases, does not guarantee that said phases can be computed efficiently in the number of compositions, nor that the resulting program could be decomposed into contiguous QSP subroutines. This work careful preserves both (1) and (2), lifting said pleasant properties to a new setting.

\subsection{Problem statement} \label{sec:problem_statement}

The definition of semantically embedded QSP depends on common constructions in category theory, covered in Appendix~\ref{sec:cat_theory}. It will turn out that the foundational category theoretic concept of \emph{natural transformations} (and its accompanying diagram) will, when assigned proper objects in quantum computation, identify criteria under which a given scheme for embedding QSP protocols becomes semantic (i.e., algorithmically meaningful). 

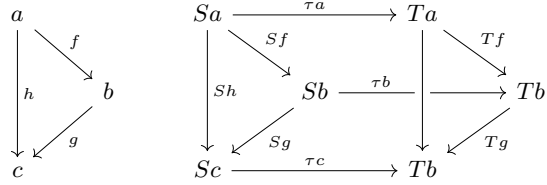
\begin{figure}[htpb]
    \centering
    \begin{tikzcd}
        a \arrow[rd, "f"] \arrow[dd, "h"] & & Sa \arrow[dd, "Sh"] \arrow[dr, "Sf"] \arrow[rr, "\tau a"] & & Ta \arrow[dr, "Tf"] \\
        & b \arrow[dl, "g"] & & Sb \arrow[dl, "Sg"] \arrow[rr, "\tau b" near start] & & Tb \arrow[dl, "Tg"] \\
        c & & Sc \arrow[rr, "\tau c"] & & Tb \arrow[from=uu, crossing over]           
    \end{tikzcd}
    \caption{A diagram presenting a natural transformation between morphisms $S$ and $T$. Special assignments from quantum algorithms to the objects depicted here provide a concrete definition for semantic embedding in Def.~\ref{def:semantic_embedding}.}
    \label{fig:qsp_as_natural_transformation}
\end{figure}

Toward our problem statement, we translate the diagram given in Fig.~\ref{fig:qsp_as_natural_transformation} (similar to Fig.~\ref{fig:natural_transformation_def} in the appendices) to the world of QSP by making a series of assignments to the objects and arrows. The basic objects, indexed by $a, b, c$ in the original figure become ordered lists of QSP phases (possibly constrained), on which the natural morphisms are functions $f, g, h$ taking two phase lists (one implicit) and embedding one inside the other to produce a new, flat list according to $\Phi_0,\, \Phi_1 \rightarrow \Phi_1 \circ \Phi_0$ with type signature $\mathbb{R}^n \times \mathbb{R}^m \rightarrow \mathbb{R}^{n\times m}$, where these functions will end up being defined to exactly correspond to the act of the outer QSP protocol (with phases $\Phi_1$) calling the inner protocol (with phases $\Phi_0$) whenever it would have called the standard QSP oracle (i.e., the QSP protocol with phases $\{0, 0\}$). This concept of physically `nesting' one protocol inside the other is vital, as in general we seek to use QSP protocols (and their functional transforms) as resources whose action can not in general be subdivided. 

We define the functors $S, T$ as the unitary resulting from compiling a QSP protocol according to a set of phases, and the result of projecting out the top-left matrix element of this unitary respectively. The latter recovers an amplitude whose magnitude-squared is a probability whose Bernoulli distribution is easy to sample from. The full map from category theoretic notation to that of QSP in Table~\ref{tab:qsp_categories}.

\begin{table}[htpb]
    \centering
    \begin{tabular}{l l l}
        Category picture\;\;\; & QSP picture & Description\\[-2.3ex]
        & &
        \\\hline
        \rule{0pt}{0.9\normalbaselineskip}$a, b, c$ & $\Phi$ & \text{QSP phase list}\\
        $f, g, h$ & $\Phi_0 \rightarrow \Phi_1 \circ \Phi_0$ & \text{QSP phase nesting (see Def.~\ref{def:nested_protocols})} \\
        $Sa, Sb, Sc$ & $U(\Phi)$ & \text{QSP unitary}\\
        $Sf, Sg, Sh$ & $U(\Phi_0) \rightarrow U(\Phi_1 \circ \Phi_0)$\;\;\;\; & \text{QSP unitary embedding}\\
        $Ta, Tb, Tc$ & $P(x)$ & \text{Polynomial transform}\\
        $Tf, Tg, Th$ & $P_0(x) \rightarrow (P_1\circ P_0)(x)$ & \text{Polynomial composition (see Thm.~\ref{thm:asym_qsp_composition})}\\
        $\tau a, \tau b, \tau c$ & $U(\Phi) \rightarrow P(x)$ & \text{Projection to matrix element}\\
    \end{tabular}
    \caption{A table relating the definition of natural transformations, and the instantiation of this diagram with objects and arrows (functions) in QSP. We suppress multi-labelling of QSP objects in the second column for brevity.}
    \label{tab:qsp_categories}
\end{table}

\begin{definition}[Semantic embedding for QSP] \label{def:semantic_embedding}
    Taking the diagram in Fig.~\ref{fig:qsp_as_natural_transformation}, we say that the morphisms $f, g, h$ are \emph{semantically embedding} for QSP protocols if $S, T$ as translated by Table~\ref{tab:qsp_categories} are such that $Sf, Sg, Sh$ correspond to circuit composition (respecting contiguity of the inner phase list), $Tf, Tg, Th$ correspond to polynomial composition, the components of the natural transformation $\tau a, \tau b, \tau c$ are simple projection from a unitary operator to its top-left matrix element, and the diagram given commutes. Moreover, the functions $f, g, h$ should be efficiently computable. $\square$
\end{definition}

In other words, we say that a method of embedding QSP protocol is \emph{semantic} if the pre-image of polynomial composition according to the functor $T$ in the space of QSP phases (i.e., $a, b, c$) is efficiently computable. While we will refer specifically to the arrows $f, g, h$ in the diagram for a natural transformation as semantic, we will sometimes overload our definitions and refer to the assignments made in Table~\ref{tab:qsp_categories} as semantic also. Note that the desirable action from a computational standpoint is to \emph{manipulate embedded polynomial transforms} of an unknown scalar (which in turn relate simply to amplitudes assigned to quantum states). Computing the preimage of this composition in the \emph{programmable space of QSP phases} makes this goal concretely satisfiable. This is equivalent to the statement of Def.~\ref{def:semantic_embedding}: that even fixing all of the natural transformation $\tau$, the functors $S, T$, and the objects $a, b, c$, still permits a satisfying (efficiently computable) definition for $f, g, h$.

\subsection{Prior work}

    The theory of QSP has its roots in the study of composite pulse techniques for NMR \cite{ylc_14, lyc_16_equiangular_gates}, and was first applied to improve methods for Hamiltonian simulation \cite{lc_17_simultation, lc_19_qubitization}. While early works considered systems of multiple qubits, the method for lifting QSP to such settings was greatly expanded to consider the manipulation of general embedded linear operators and renamed QSVT \cite{gslw_19}. The core of this argument concerned the QSP-like manipulation of invariant SU(2) subspaces preserved by alternating projectors according to an old result: Jordan's lemma \cite{jordan_75}. Recent work has also simplified the application and interpretation of Jordan's lemma in QSVT, relying on the simpler but related cosine-sine decomposition \cite{cs_qsvt_tang_tian}.
        
    QSP and QSVT have since been applied to numerous disparate problems in quantum algorithms: Hamiltonian simulation \cite{coherent_ham_sim_21}, phase estimation \cite{rall_21}, quantum zero-knowledge proofs \cite{lombardi_pqzk_2021}, classical quantum-inspired machine learning \cite{chia_20}, semi-definite programming \cite{q_sdp_solvers_20}, quantum adiabatic methods \cite{lin_eig_filter_20}, computation of approximate correlation functions \cite{rall_correlation_20}, computation of approximate fidelity \cite{gilyen_fidelity_22}, recovery maps \cite{petz_recovery_20}, metrology and calibration \cite{dgn_qsp_metrology_22}, and fast inversion of linear systems \cite{tong_inversion_21}. Such breadth has earned QSVT a qualified description as a unifying quantum algorithm \cite{mrtc_21}.
    
    Simultaneously, detailed work has been done to ensure that the classical algorithms for determining the parameterizations of QSP and QSVT circuits are truly efficient and numerically stable. These include initial work on using standard precision arithmetic \cite{chao_machine_prec_20, haah_2019, dong_efficient_phases_21}, as well as methods for removing unnecessary degrees of freedom in the QSP circuit parameterization \cite{sym_qsp_21}. Recent work has also further clarified the properties of the map between QSP phases and the achieved functional transform in terms of the norm of the former and the Fourier coefficients of the latter \cite{dlnw_infinite_qsp_22}.
    
    Research into the coherent composition and combination of quantum subroutines has been more limited, in part due to the difficulty of analyzing their behavior beyond those protocols simply using amplitude amplification. These simpler instances include quantum routines for scheduling \cite{grover_scheduling_02} (which predates and thus does not use QSP explicitly), and distributed linear algebraic problems \cite{ms_communication_complexity_22}. Additionally and excitingly, the black-box use of quantum subroutines is prevalent in proofs of security for quantum cryptography, wherein quantum analogues for common classical cryptographic techniques for rewinding (e.g., witness extraction and simulation) \cite{lombardi_pqzk_2021, cmsz_qpsa_2022} often require one to precisely understand the properties of such coherent circuit compositions. Even beyond the cryptographic setting, the study of quantum superchannels with black-box access is rich with applications to quantum channel discrimination and the communication of quantum information \cite{acin_channel_disc_01, harrow_adaptive_disc_10, duan_feng_ying_disc_07, duan_perfect_disc_09, takagi_channel_disc_19, ysm_channel_inverse_22}.

    This work combines elements from all of the above-mentioned areas: (1) basic considerations on the theory of QSP/QSVT, (2) properties of QSP/QSVT protocols with restricted phases, and (3) the recent tools available for the analysis of distributed or multi-party quantum computations. Finally, we employ basic concepts and language from the world of category theory, which has had success, as inspired by its classical counterpart in programming language design and program verification \cite{hoare_axiomatic_69, floyd_program_meaning_93, apt_verification_09}, in undergirding a growing theory of quantum programming languages. A long line of work has considered such category-theoretical constructions of higher-level quantum programming concepts \cite{selinger_qpl_04, ac_category_semantics_04}, including data-types capturing purity in quantum computations \cite{yuan_twist_22}, interpretations for recursion and self-embedding \cite{xyv_recursive_programs_21, ying_control_flow_12}, and methods for verifying program correctness based in Floyd-Hoare logic \cite{ying_floyd_hoare_12}. Determining if QSVT can contribute meaningfully to such abstractions is one focus of this work.

\subsection{Summary of results and open problems}

    This work is divided into four major parts: (1) a problem statement and definition for \emph{semantic embedding} based on concepts in category theory, discussed in Sec.~\ref{sec:problem_statement}, (2) a series of theorems constructing a satisfying assignment for this definition using QSP in Sec.~\ref{sec:semantic_embedding_qsp}, (3) a series of theorems discussing two different instances of a satisfying assignment for this definition using QSVT in Sec.~\ref{sec:semantic_embedding_qsvt}, and finally (4) a series of expanded examples of known quantum algorithms which implicitly make use of semantic embedding in Sec.~\ref{sec:applications}, from a variety of disparate subfields. Where possible we discuss where our constructions are unique, and in what sense the exhaust the possible satisfying assignments to the diagram for semantic embedding. This work is targeted toward an audience familiar with the theory of QSP and QSVT, and as such basic results and properties of these algorithms are left, for the interested reader, to Appendix~\ref{sec:qsp_qsvt_theory}.

    The new components in this work exist on two levels. On the first, we provide a succinct explanation, rooted in category theory, for what a variety of nested QSP and QSVT circuits are achieving in terms of their underlying embedded transformations. We show that this diagrammatic interpretation leads to greater clarity for what is both desirable and possible to achieve by the coherent composition of QSP/QSVT as subroutines. Secondly, we work through the details of nested QSP and QSVT protocols to determine the concrete constraints on their circuit parameters that enable not merely \emph{nested} protocols, but \emph{embedded} functional transforms. Throughout this work, we try to use these two terms strictly: \emph{nesting} is a process in the programmable space of the circuit ansatz (e.g., the QSP phases $\Phi$) that respects blocks of quantum gates as subroutines, while \emph{embedding} is a process in functional space (e.g., the polynomial transform $P(x)$ induced by a QSP unitary). The goal for the embedding is for it to be semantic---logical and meaningful---while also being induced (in circuit space, and up to mild constraints) by nesting. Definitions and theorems in Sec.~\ref{sec:semantic_embedding_qsp} and Sec.~\ref{sec:semantic_embedding_qsvt} are structured in parallel to this distinction.

    Open problems include expanding the possible satisfying assignments for the functors in Fig.~\ref{fig:qsp_as_natural_transformation} in the setting of QSP, and more strongly characterizing possible manipulations in functional space. Moreover, as is true for QSVT generally, most applications basically perform amplitude amplification, which is often achievable nearly as efficiently through other means. Determining interactive protocols (perhaps like those in \cite{ms_communication_complexity_22}) for which round and communication complexity are both non constant may help in this goal. Finally, this work applies only extremely basic category theoretic concepts; formalizing useful abstractions in quantum computing for the manipulation of quantum information using QSP/QSVT is still nascent, and may have great use in the development of higher-level quantum programming languages as we attempt to move away from a gate-level understanding of quantum computations.

\section{Semantically embedded QSP} \label{sec:semantic_embedding_qsp}

Toward a satisfying assignment for semantic embedding (Def.~\ref{def:semantic_embedding}), we use this section to define and discuss the result of successively nesting QSP protocols (Def.~\ref{def:nested_protocols}), or equivalently the action of QSP protocols which call, in place of their standard oracle, the result of another, consistent, perhaps unknown QSP computation. We are able to show that only mild restrictions on these protocols lead to a satisfying assignment for the $f, g, h$ given in the diagram of Fig.~\ref{fig:qsp_as_natural_transformation}, and prove a variety of key properties about the achievable functional transforms.

\begin{figure}
    \centering
    \includegraphics[width=0.9\textwidth]{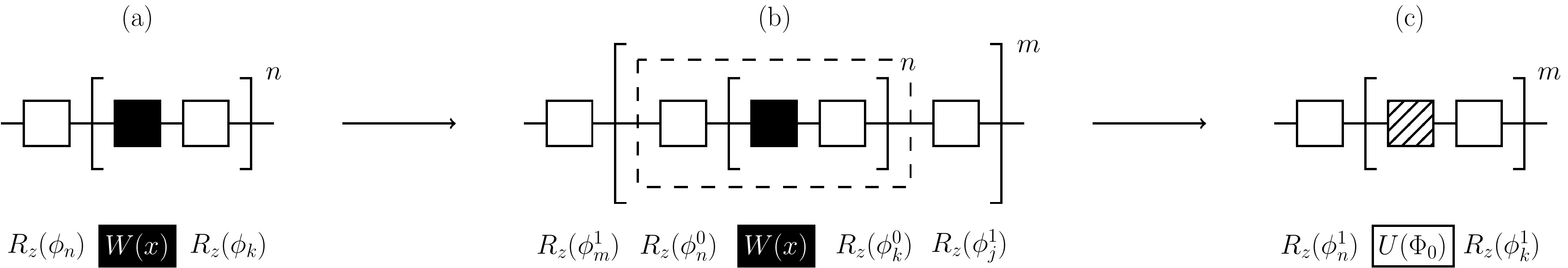}
    \caption{Symbolic representation for the nesting of QSP protocols (Def.~\ref{def:nested_protocols}). Here (a) shows standard QSP, an interleaved series of control (white) and oracle (black) unitaries, (b) shows one instance of a QSP protocol nested in another, while (c) gives the same, with the nested protocol (diagonal hashes) in a suppressed notation (i.e., treated as an oracle). This supressed notation becomes more useful in the depiction of self-embedded QSVT in Fig.~\ref{fig:qsvt_composition}. Here $R_z(\phi) = \exp{(i\,\phi\,\sigma_z)}$ while $W(x) = \exp{(i\arccos{x}\,\sigma_x)}$ per usual, and $U(\Phi)$ is the QSP protocol with phases $\Phi$.}
    \label{fig:qsp_composition}
\end{figure}

\begin{definition}[Nested QSP protocol] \label{def:nested_protocols}
    A nested QSP protocol is identical to a standard QSP protocol up to the substitution of the standard oracle by \emph{another QSP protocol}. I.e., given $U_{\Phi_0}(x)$, the unitary generated by a QSP protocol with QSP phases $\Phi_0 \in \mathbb{R}^{n + 1}$, the (once) nested QSP protocol according to phase lists $\Phi_1 \in \mathbb{R}^{m + 1}, \Phi_0 \in \mathbb{R}^{n + 1}$ has the defining quantum circuit:
        \begin{equation} \label{eq:circuit_comp}
            (\Phi_1 \circ \Phi_0)(x) =
            e^{i\phi_{1, 0}}\prod_{k = 1}^{m} U_{\Phi_0}(x) e^{i\phi_{1, k} \sigma_z},
        \end{equation}
    where $\phi_{1, k}$ is the $k$-th element of $\Phi_1$ and analogously for $\Phi_0$. We use $(\Phi)(x)$ as shorthand for $U_{\Phi}(x)$ in a way that makes composition of such protocols, e.g., $(\Phi_1 \circ \Phi)(x)$ easy to express. Here $\Phi_0$ specifies the \emph{inner protocol} while $\Phi_1$ specifies the \emph{outer protocol}. This nesting can be continued recursively as many times as one wishes, e.g.,
        \begin{equation}
            (\Phi_{n} \circ \cdots \circ \Phi_1 \circ \Phi_0)(x).
        \end{equation}
    A depiction of these protocols is given in Fig.~\ref{fig:qsp_composition}. Note that the inner protocol's phases remain contiguous under nesting, implying that this protocol is really being used as a black box. $\square$
\end{definition}

We can see clearly that nested QSP protocols are, by simple expansion, valid QSP protocols, but that the composition of two arbitrary protocols does not compose their functional transforms. A simple example of this follows taking $\Phi_0 = \Phi_1 = \{\pi/4, \pi/4, \pi/4\}$, whose inner and nested protocols embed the following respectively
    \begin{align}
        P &= -(1 + i)/\sqrt{2} + i\sqrt{2} x^2,\\
        P^\prime &= -(1 + i)/\sqrt{2} - (1 - i) \sqrt{2} x^2 + 2 \sqrt{2} x^4,
    \end{align}
where the latter is clearly different than the composition of $P$ with itself:
    \begin{equation}
        (P\circ P) = -(3 + i)/\sqrt{2} +  (1 + i) 2\sqrt{2} x^2 - i 2 \sqrt{2} x^4 \neq P^\prime.
    \end{equation}
One can also see that taking instead $\Phi_0 = \Phi_1 = \{0, 0, 0\}$ does indeed induce the composition of embedded polynomial transforms $T_2(x) \mapsto (T_2 \circ T_2)(x) = T_4(x)$ where $T_n(x)$ is the $n$-th Chebyshev-$T$ polynomial evaluated at $x$. The primary aim of a theory of semantically embedded QSP protocols is thus to formally specify the conditions under which some reasonable form of nesting for circuits (a manipulation in the programmable space preserving contiguousness of the inner protocol), corresponds neatly to polynomial composition (or a related manipulation in functional space).

While the counterexample given above for general composition of functional transforms in QSP seems to rely heavily on properties of SU(2), the same sorts of counterexamples can be made to appear in the theory of classical Boolean functions. We give a short exposition of one such counterexample, illustrating that the composition of general functions in even classical data processing ought to be treated carefully.

\begin{example}[Composing multivariable Boolean functions] \label{ex:boolean_composition}
    Consider two multivariable Boolean functions $f, g$ with the same domain and range:
        \begin{equation}
            f, g : \{0, 1\}^4 \rightarrow \{0, 1\}^2.
        \end{equation}
    For purpose of example we define their action the following way:
        \begin{align}
            f(x_0, x_1, x_2, x_3) &= \{x_0 \land x_1, x_2\land x_3\}\\
            g(x_0, x_1, x_2, x_3) &= \{x_0 \lor x_1, x_2\lor x_3\}.
        \end{align}
    We note that if one wanted to compute the logical and ($\land$) or logical or ($\lor$) of two variables, they could use one of the functions above, ignoring select bits from the input and output, i.e.,
        \begin{equation}
            x_0 \land x_1 = f^{01\mapsto 0},
        \end{equation}
    where by this notation we mean that we care only about what we feed to the $0,1$ indices of $f$, and what is output in the $0$ index. Now consider the following `natural' composition rule (i.e., appending the outputs of two applications of $k$ and feeding this into $h$) for functions $h, k$ with type signature $\{0, 1\}^4 \rightarrow \{0, 1\}^2$, namely
        \begin{align}
            h\circ k &: \{0, 1\}^4 \rightarrow \{0, 1\}^2\\
            (h\circ k)(x) &= h(k^{0123\mapsto 0}(x),k^{0123\mapsto 1}(x),k^{0123\mapsto 0}(x),k^{0123\mapsto 1}(x)),
        \end{align}
    where we have condensed the four elements of the argument into the tuple $x$. If one is now to compute $(h\circ k)^{01\mapsto 0}$, it is clearly not the `natural' composition of the function $h^{01\mapsto 0}$ with that of $k^{01\mapsto 0}$, each of which have type signatures $\{0, 1\}^2 \rightarrow \{0, 1\}^1$. For our functions $f, g$ the composition according to this rule results in
        \begin{equation}
            (g \circ f)^{01\mapsto 0} = (x_0\land x_1)\lor(x_2 \land x_3),
        \end{equation}
    rather than the intended composition
        \begin{equation}
            (g \circ f)^{01\mapsto 0} = (x_0\land x_1)\land(x_0 \land x_1),
        \end{equation}
    demonstrating our intended result. $\square$
\end{example}

While the functions given in Example~\ref{ex:boolean_composition} are somewhat contrived, the composition rule provided is not unnatural at first glance, and can in fact be simply repaired if the rule for composition is replaced with a permuted version of itself, namely
    \begin{align}
        (h\circ k)^\prime &: \{0, 1\}^4 \rightarrow \{0, 1\}^2\\
        (h\circ k)^\prime(x) &= (h\circ \pi k)\\
        &= h(k^{0123\mapsto 0}(x),k^{0123\mapsto 0}(x),k^{0123\mapsto 1}(x),k^{0123\mapsto 1}(x)),
    \end{align}
where here $\pi$ is some permutation swapping the second and third outputs of the function $k$. The key takeaway is that when the underlying accessible maps are situated in a larger natural domain and range than the one intended for evaluation (as is the case for the unitary $U_\Phi$ and the physically accessible probability $P(x)$ in QSP), then composition of these underlying desirable functions (e.g., $f^{01\mapsto 0}$ and $g^{01\mapsto 0}$) is not only not assured, but often \emph{not possible} without access to additional manipulations. In this work we consider quantum protocols with black-box access to QSP/QSVT protocols, and consequently being able to flexibly make use of these subroutines in computations carries the caveat that the required manipulations to respect functional manipulations embedded in these protocols need to be physically possible to perform without explicit knowledge of the oracular protocol. This in turn means we have to explicitly consider the action of these functions in domains larger than those required for computation, as in the above example.

Before discussing a satisfying assignment for the question posed in Def.~\ref{def:semantic_embedding}, we give a definition for QSP protocols with a property sufficient to avoid the problem discussed in the previous paragraph. I.e., we define protocols which, upon nesting (Def.~\ref{def:nested_protocols}) induce functional composition of their top-left elements. Throughout this work will consider the top-left element for clarity, but we could have privileged other measurement bases as well with similar results.

\begin{definition}[Embeddable QSP protocol] \label{def:embeddable_protocols}
    Let $U_\Phi(x)$ be the unitary for a QSP protocol with signal $x$ and phases $\Phi$. A QSP protocol is \emph{embeddable} if for all $x \in [-1, 1]$ the unitary $U_\Phi(x)$ has the form
        \begin{equation} \label{eq:embeddable_qsp}
            U_\Phi(x) = \exp\biggr[\,i\big\{f(x)\,\sigma_x + g(x)\,\sigma_y\big\}\biggr],
        \end{equation}
    where $f, g: [-1, 1] \mapsto \mathbb{R}$ are functions of $x$ where $|f(x)|^2 + |g(x)|^2 = 1$ for all $x \in [-1,1]$ and $f(\pm 1) = g(\pm 1) = 0$. Equivalently, embeddable QSP protocols are those for which the resulting unitary, for any signal $x$, rotates about an axis in the $XY$-plane of the Bloch sphere. $\square$
\end{definition}

To see that QSP protocols satisfying Def.~\ref{def:embeddable_protocols} can be nested to compose their embedded polynomial transforms, it is enough to observe the following lemma.

\begin{lemma}[QSP with twisted oracles] \label{lemma:twisted_qsp}
    Let $W(x) = e^{i\arccos{x}}$ the standard QSP oracle. Then the \emph{twisted oracle} $W_{\theta}(x) = e^{i\theta\sigma_z}W(x)e^{-i\theta\sigma_z}$ can be used as the oracle for a standard QSP protocol obliviously, and produces a \emph{twisted unitary transform} that relates to the untwisted unitary transform simply:
        \begin{equation}
            U_{\Phi}(\theta, x) = e^{i\theta\sigma_z}U_{\Phi}(x)e^{-i\theta\sigma_z},
        \end{equation}
    where $U_{\Phi}(\theta, x)$ is $U_{\phi}(x)$ using twisted oracles in place of untwisted ones. Proof is easily seen by expanding the definition of QSP protocols and cancelling adjacent opposite $\sigma_z$ rotations (save those at the beginning and end). $\square$
\end{lemma}

In other words, the form given in Def.~\ref{def:embeddable_protocols} is precisely that of a twisted oracle, meaning that fully characterizing the properties of QSP phase lists which produce embeddable protocols will in turn characterize the properties required for semantic embedding in QSP (Def.~\ref{def:semantic_embedding}). We start this by defining a sub-class of QSP protocols whose phases obey a discrete symmetry. The study of such restrictions on QSP phases has recently expanded in scope, and we give a view of these advances in Appendix~\ref{sec:constrained_qsp_phases}.

\begin{definition}[(Even) antisymmetric QSP protocols] \label{def:asym_qsp_protocol}
    Let $\Phi \in \mathbb{R}^{2d}$; a length $2d$ antisymmetric QSP protocol is a QSP protocol with $\Phi = \Phi^\prime \cup (\Phi^{\prime})^{RN}$ for $\Phi^{\prime} \in \mathbb{R}^{d}$ arbitrary. Equivalently, $\Phi$ is antisymmetric under phase-order reversal:
        \begin{equation}
            \Phi = \{\phi_0, \phi_1, \phi_2, \cdots, \phi_d, -\phi_d, \cdots, -\phi_2, -\phi_1, -\phi_0\}.
        \end{equation}
    Note that this constrains $P$ to be real, as discussed in Appendix~\ref{sec:constrained_qsp_phases}. We can also define odd antisymmetric protocols with $\Phi \in \mathbb{R}^{2d + 1}$:
        \begin{equation}
            \Phi = \{\phi_0, \phi_1, \phi_2, \cdots,\phi_{d}, 0,-\phi_{d}, \cdots, -\phi_2, -\phi_1, -\phi_0\},
        \end{equation}
    where the central phase is constrained to be zero. $\square$
\end{definition}

Antisymmetric QSP protocols, by merit of the mild symmetry imposed on their defining phase lists, are still quite expressive. In what follows, we prove a variety of their most salient properties, before connecting them to embeddable QSP protocols discussed earlier. We express that studying the properties of QSP protocols whose phases obey constraints is an interesting direction in itself, with bearing on numerical properties of algorithms optimizing over such protocols \cite{sym_qsp_21}, as well as their action under noise \cite{tltc_alec_23}.

\begin{theorem}[Existence and uniqueness of antisymmetric QSP protocols] \label{thm:existence_uniqueness_asym_qsp}
    Take polynomials $P \in \mathbb{R}[x]$ (note real) and $Q \in \mathbb{C}[x]$ satisfying the following:
        \begin{enumerate}
            \item The degree of $P$ is $d$ and the degree of $Q$ is $d - 1$.
            \item $P$ has parity $d \pmod 2$ and $Q$ has parity $(d - 1) \pmod 2$.
            \item $\forall x \in [-1,1]$, $P, Q$ satisfy $|P|^2 + (1 - x^2)|Q|^2 = 1$.
            \item In the case that $d$ is even, the leading coefficient of $P$ is positive.
        \end{enumerate}
    There exists a unique set of antisymmetric phase factors $\Phi = \{\phi_0, \phi_1, \cdots, -\phi_1, -\phi_0\} \in D_{d}$ such that
        \begin{equation} \label{eq:asym_explicit_form}
            e^{i\phi_0\sigma_z}\prod_{k = 1}^{d - 1} W(x)e^{i\phi_k\sigma_z}
            =
            \begin{bmatrix}
                P & Q\sqrt{1 - x^2}\\
                -Q^{*}\sqrt{1 - x^2} & P
            \end{bmatrix}.
        \end{equation}
    Here $D_d$, the antisymmetric QSP phase domain, following the notation of \cite{sym_qsp_21}, and is defined in the even and odd cases (Def.~\ref{def:asym_qsp_protocol}) separately:
        \begin{alignat}{3}
            D_{d} &= [-\pi/2, \pi/2)^{2d}\quad\quad &&\text{for even protocols},\label{asym_qsp_domain_1}\\
            D_{d} &= [-\pi/2, \pi/2)^{d}\times\{0\}\times[-\pi/2, \pi/2)^{d}\quad\quad &&\text{for odd protocols}.\label{asym_qsp_domain_2}
        \end{alignat}
    Note in our theorem statement both $P, Q$ are specified; if only $P$ is specified, then multiple satisfying $Q^\prime$, e.g., $Q^*$ can be chosen satisfying $|P|^2 + (1 - x^2)|Q|^2 = 1$, each of which may yield different (antisymmetric) phases. However, if one restricts the roots of $Q(z + 1/z)$ in $z$ to lie either entirely inside or entirely outside the unit circle in $\mathbb{C}$, then this choice can be made unique for a given $P$ \cite{rossi_m_qsp_22, haah_2019}. For proof see Appendix~\ref{sec:embedded_qsp_qsvt}. $\square$
\end{theorem}

\begin{theorem}[Antisymmetric and embeddable QSP protocols.] \label{thm:asym_embedded_qsp_equivalence}
    Antisymmetric QSP protocols are in bijection with embeddable QSP protocols under phase domain restriction to $D_d$, and under the assumption that $f, g$ as in Def.~\ref{def:nested_protocols} are polynomial functions of $x$. For proof see Appendix~\ref{sec:embedded_qsp_qsvt}. $\square$
\end{theorem}

Additionally, we provide an analogue of the main matrix-completion theorem of QSP (see Appendix~\ref{sec:qsp_qsvt_theory}) for antisymmetric QSP protocols. This constraint, as shown, adds a new condition on the roots of the embedded polynomials, which is easy to describe. In practice, one finds QSP phases for such protocols with numerical optimization, but it is an interesting question whether there exist simpler, alternative methods of describing polynomials with similar properties of their root sets.

\begin{theorem}[Partially specified antisymmetric QSP protocols] \label{thm:partial_asym_qsp}
    Let $P \in \mathbb{R}[x]$ (note real) of degree $d$ satisfying the following conditions:
        \begin{enumerate}
            \item $P$ has parity $d \pmod 2$.
            \item $\forall x \in [-1, 1]$, $|P(x)| \leq 1$.
            \item $|P(\pm 1)| = 1$.
            \item For the following expression:
                \begin{equation}
                    1 - P([z + z^{-1}]/2)^2 = F(z) = \alpha\prod_{r_i} (z - r_i)(z^{-1} - r_i),
                \end{equation}
                the multiset $r_i$ of roots of $F$ must be closed under negation.
            \item In the case that $d$ is even, the leading coefficient of $P$ is positive.
        \end{enumerate}
    There exists a unique antisymmetric QSP protocol with $\Phi \in D_d$ (Eqs.~\ref{asym_qsp_domain_1}, \ref{asym_qsp_domain_2}) whose unitary has the form
        \begin{equation}
            U_\Phi(x) =
            \begin{bmatrix}
                P & Q\sqrt{1 - x^2}\\
                -Q^{*}\sqrt{1 - x^2} & P
            \end{bmatrix}.
        \end{equation}
    Moreover, the phases $\Phi$ can be efficiently computed classically given the coefficients of $P$.
    \begin{proof}
        Proof follows by standard application of the single-variable Fejér-Riesz lemma \cite{rossi_m_qsp_22}. Take $A(x) = 1 - P(x)^2$, evidently a non-negative polynomial with roots at $x = \pm 1$. The nonnegativity of $A(x)$ implies that it is expressible as the modulus squared of a complex polynomial $R(x)$ with a prefactor, i.e,
            \begin{equation}
                1 - P(x)^2 = A(x) = (1 - x^2)|R(x)|^2,
            \end{equation}
        where we have factored out the known roots at $x = \pm 1$. Taking $\mathfrak{R}(Q) = \mathfrak{R}(R)$ and $\mathfrak{I}(Q) = \mathfrak{I}(R)$ recovers the desired relation $|P(x)|^2 + (1 - x^2)|Q(x)|^2 = 1$. By the uniqueness of antisymmetric QSP protocols (Thm~\ref{thm:existence_uniqueness_asym_qsp}), this completion is also unique.
    \end{proof}
\end{theorem}

\begin{theorem}[Nesting of antisymmetric/embeddable QSP protocols] \label{thm:asym_qsp_composition}
    Let $\Phi_0, \Phi_1$ be the (unique) antisymmetric (equivalently embeddable) QSP phase lists embedding polynomial transforms $P_0, P_1$ in their corresponding unitaries. Equivalently,
        \begin{equation}
            P_0 = \langle 0 | U_{\Phi_0}(x)|0\rangle, 
            \quad 
            P_1 = \langle 0 | U_{\Phi_1}(x)|0\rangle.
        \end{equation}
    The nesting of these QSP protocols results in the following unitary:
        \begin{equation}
            (\Phi_1 \circ \Phi_0)(x) = 
            \begin{bmatrix}
                (P_1 \circ P_0)(x) & \cdot \\
                \cdot & (P_1 \circ P_0)(x)
            \end{bmatrix},
        \end{equation}
    where the off-diagonal elements are easily computed using Thm.~\ref{thm:partial_asym_qsp} if desired. In other words \emph{nesting of embeddable QSP protocols} implies a corresponding \emph{composition of polynomial transforms}. Moreover, as a corollary, we see that both the embeddable and asymmetric properties of QSP protocols are preserved under composition.
    
    \begin{proof}
        Proof follows easily from application of Lemma~\ref{lemma:twisted_qsp}, recognizing the inner QSP protocol, $\Phi_0$, as producing a twisted oracle encoding $P_0$ (as seen by $\Phi_1$. Consequently, up to an overall $\sigma_z$-rotation, $P_1$ and $P_0$ compose, and moreover the resulting protocol, by the fact that both $P_0, P_1$ are real-valued, is also embeddable and thus antisymmetric.
    \end{proof}
\end{theorem}

\begin{corollary}[Uniqueness of antisymmetric phase factors for nested QSP and polynomial composition] \label{cor:uniqueness_for_antisymmetry}
    Given an assignment for the diagram given in Fig.~\ref{fig:qsp_as_natural_transformation} where the arrows $f, g, h$ correspond to the outer QSP protocol calling the inner QSP protocol in place of its standard oracle, antisymmetry of the phase lists $\Phi$ as defined in Def.~\ref{def:asym_qsp_protocol} is the unique symmetry which for all $x$ permits this diagram to commute.
    \begin{proof}
        Proof of this fact is relatively straightforward. Take a generic QSP protocol whose phases do not obey the antisymmetry constraint given in Def.~\ref{def:asym_qsp_protocol}. By definition this protocol embeds in its corresponding unitary's top left corner a polynomial $P$ which is not entirely real on its domain, as otherwise there would exist an equivalent set of phases which both achieved this polynomial transform and were antisymmetric. Consequently this oracle is not a rotation about an axis in the XY plane of the Bloch sphere for all $x$, and thus is not generated by an element of SU(2) which anticommutes with Z rotations for all arguments $x$. Therefore there exists a non-trivial (i.e., non-constant) component of the oracle QSP protocol which commutes with all applied $Z$ rotations for $x \neq \pm 1$, and can therefore not be involved in the functional composition. But we know a general outer protocol can provide a $P$ which is injective its domain, meaning that our intended functional composition has necessarily lost information about the action of the inner protocol (i.e., the commuting part of the oracle).
    \end{proof}
\end{corollary}

Returning to the original statement of semantic embedding (Def.~\ref{def:semantic_embedding}), and following the prescription of Table~\ref{tab:qsp_categories}, we see that restricting to antisymmetric QSP phase lists (objects $a, b, c$ in Fig.~\ref{fig:qsp_as_natural_transformation}), permits nesting of these phase lists (arrows $f, g, h$ defined according to Def.~\ref{def:nested_protocols}) to correspond directly (through the components of the natural transformation $\tau a, \tau b, \tau c$) to functional composition in the picture provided by the functor $T$. Moreover, due to the intimate relation between the condition imposed by antisymmetry, and properties of the Lie group SU(2), we can make a strong statement for the uniqueness of this constraint for achieving functional composition (on all arguments and exactly) by composing circuits (Corollary~\ref{cor:uniqueness_for_antisymmetry}). Consequently, restriction to antisymmetric phase lists (which as shown above imposes only mild restrictions on the achievable embedded polynomial transformations), permits the diagram depicting the natural transformation (with assignments to the objects and arrows of QSP) to commute. In what follows, we will take this satisfying assignment and lift the resulting QSP protocols to QSVT. We note that it is an interesting and open question if such constraints can be relaxed if one imposes additional constraints on the magnitude of the involved signals, or if one desires composition to only be approximately achieved, but we leave this discussion to future work.

\section{Semantically embedded QSVT} \label{sec:semantic_embedding_qsvt}

The aims of this section are twofold. The first is to follow the standard lifting argument from QSP to QSVT, covered in Appendix~\ref{sec:lifting_qsp_qsvt}, to give a theory of semantic embedding for QSVT. The second is to investigate additional freedoms in the structure of QSVT to devise new satisfying assignments for the diagram in Fig.~\ref{fig:qsp_as_natural_transformation} which have distinct interpretations in both the programmable and functional spaces of QSVT. In turn, we want to argue that these expanded satisfying assignments, together with those inherited directly from QSP, in some strong sense constitute all reasonable satisfying assignments; toward this, we consider the closure of quantum circuits (and their induced transforms) under such manipulations.

The first of these nesting procedures follows almost immediately from the QSP nesting procedure in Def.~\ref{def:nested_protocols}, respecting the order of application of oracles in the QSP case, and consequently inducing an identical nesting transformation in each two-dimensional subspace preserved by the actions of the two projectors $\Pi, \tilde{\Pi}$ constituting the relevant QSVT protocol. These projectors are in some sense the main distinguishing factor between QSP and QSVT, locating the sub-block of a unitary matrix to whose eigenvalues or singular values our QSP-like polynomial transformations are applied \cite{gslw_19, cs_qsvt_tang_tian}. In this way we can, in analogy to the type signature for QSP in Eq.~\ref{eq:qsp_type_signature}, we can give an analogous one for QSVT including the relevant projectors
    \begin{equation} \label{eq:qsvt_type_signature}
        \text{QSVT}: \mathbb{R}^{n+1} \times \mathbb{C}^{p\times q} \times \mathbb{C}^{q\times r} \rightarrow \mathbb{C}_n[x],
    \end{equation}
where the second and third arguments will be the orthogonal projectors $\tilde{\Pi}$ and $\Pi$ in QSVT. Here the implicit scalar argument $x$ now stands for each singular value of the embedded operator $A = \tilde{\Pi}U\Pi \in \mathbb{C}^{p\times r}$ induced by the choice of projectors and the block encoding unitary $U \in \mathbb{C}^{q\times q}$. Both $U$ and $U^\dagger$ are assumed to be given to the computing party as oracles. In this case, the relevant polynomial is applied uniformly, just as it was in QSP, within subspaces defined by the left and right singular vectors of $A$; for further details on this map, see Appendix~\ref{sec:qsp_qsvt_theory}.

\begin{definition}[Flatly nested QSVT] \label{def:qsvt_flat_nesting}
    Let $\Phi_0 \in \mathbb{R}^{n}$ and $\Phi_1 \in \mathbb{R}^m$ define two QSVT protocols according to the definition in Appendix~\ref{sec:qsp_qsvt_theory}, i.e., the two descriptions of superoperators
        \begin{equation}
            (\Phi_0, \tilde{\Pi}, \Pi, \ast), (\Phi_1, \tilde{\Pi}, \Pi, \ast),
        \end{equation}
    where we have followed the condensed notation defined in Appendix~\ref{sec:condensed_notation_qsp_qsvt}. Then the flat nesting of the protocol defined by $\Phi_0$ into that defined by $\Phi_1$ is the following description of a superoperator:
        \begin{equation} \label{eq:flat_embedding}
            (\Phi_1 \circ \Phi_0, \tilde{\Pi}, \Pi, \ast),
        \end{equation}
    where $(\Phi_1 \circ \Phi_0)$ is the same nesting of $\Phi_0$ into $\Phi_1$ given in Def.~\ref{def:nested_protocols}. In this case, the action in each two-dimensional invariant subspace preserved by $\tilde{\Pi}, \Pi$ is the composition of $F, G$ as achieved by $\Phi_0, \Phi_1$ respectively. Note that both the inner protocol and outer protocol share identical subspaces for each of their basic QSVT components. $\square$
\end{definition}

The more involved definition of QSVT permits us to modify the method of nesting two protocols from merely that which respects the QSP case under the standard lifting argument. In what follows we specify one of these nesting protocols (termed deep nesting), and in the following section (Sec.~\ref{sec:applications}) show that it appears often in many, seemingly unrelated quantum algorithms. The primary difference between flat and deep nesting will be that while previously the two block-encoded operators were with respect to the same projectors, this need not in general be the case, and when the projectors are particularly simple, the interpretation of the resulting nested circuit can be made to satisfy the requirements of the diagram in Fig.~\ref{fig:qsp_as_natural_transformation} in a unique way: functional composition becomes a functional product.

\begin{definition}[Deeply nested QSVT] \label{def:qsvt_deep_nesting}
    Recall that QSVT is usually presented as a product of interleaved iterates (here for $n$ even):
        \begin{equation} \label{eq:qsvt_original_circuit}
            U_{\Phi_0} = 
            e^{i\phi^0_{1}(2\tilde{\Pi}_0 - I)}U_0
            \prod_{j = 1}^{(n - 1)/2}
            \left(e^{i\phi^0_{2j}(2\Pi_0 - I)} U_0^{\dagger}e^{i\phi^0_{2j + 1}(2\tilde{\Pi}_0 - I)}U_0\right).
        \end{equation}
    If we introduce, instead of the projector $\tilde{\Pi}_0$, a transformed projector
        \begin{equation} \label{eq:projector_transform}
            \tilde{\Pi}_0 \mapsto U_{\Phi_1}^{\vphantom{\dagger}} \tilde{\Pi}_0 U_{\Phi_1}^{\dagger},
        \end{equation}
    for some additional QSVT protocol $U_{\Phi_1}$, then the circuit in Eq.~\ref{eq:qsvt_original_circuit} can be transformed according to
        \begin{equation} \label{eq:qsvt_full_form}
            (\Phi_1 \land \Phi_0) =
            U_{\Phi_1}e^{i\phi^0_{1}(2\tilde{\Pi}_0 - I)}U_{\Phi_1}^{\dagger}U_0
            \prod_{j = 1}^{(n - 1)/2}
            \left(e^{i\phi^0_{2j}(2\Pi_0 - I)} U^{\dagger}_0 U_{\Phi_1}e^{i\phi^0_{2j + 1}(2\tilde{\Pi}_0 - I)}U_{\Phi_1}^{\dagger}U_0\right).
        \end{equation}
    We call $(\Phi_1 \land \Phi_0)$ the QSVT circuit with that enacts a deep nesting by consistently conjugating of one of the \emph{inner protocol's projectors} (here $\tilde{\Pi}_0$) by another QSVT protocol defined by phases $\Phi_1$. In this case the transformation refers equivalently to the left-multiplication of $U$ by $U_{\Phi_1}^\dagger$:
        \begin{equation} \label{eq:block_encoding_transform}
            U_0 \mapsto U_{\Phi_1}^{\dagger }U_0
            \quad\text{equiv.}\quad
            U_0^{\dagger} \mapsto U_0^{\dagger}U_{\Phi_1},
        \end{equation}
    and consequently the encoded linear operator $A = \tilde{\Pi}_0 U_0 \Pi_0$ is modified to $A_1 = \tilde{\Pi}_0 U_{\Phi_1}^\dagger U_0 \Pi_0$. $\square$
\end{definition}

\begin{figure}
    \centering
    \includegraphics[width=\textwidth]{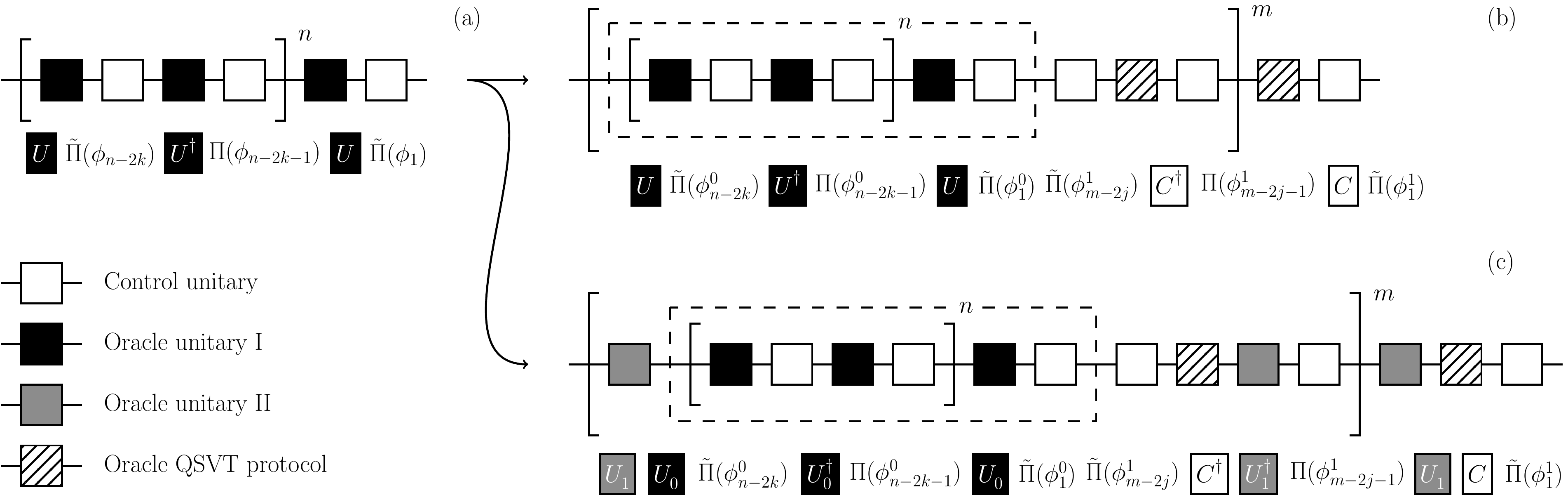}
    \caption{Symbolic representation for the composition of QSVT protocols (Defs.~\ref{def:qsvt_flat_nesting} and \ref{def:qsvt_deep_nesting}). Here (a) indicates standard QSVT, while (b) shows flat embedding (Def.~\ref{def:qsvt_flat_nesting}) and (c) shows deep embedding (Def.~\ref{def:qsvt_deep_nesting}). In deep embedding, there are two implicit unitary oracles, while the same is not true of flat embedding, which resembles the self-embedding of QSP (Def.~\ref{def:nested_protocols}). Diagonally-hashed boxes are used as in Fig.~\ref{fig:qsp_composition} to suppress notation for an oracular QSVT protocol. Here $U$ is the unitary containing a block encoding, while $\Pi(\phi)$ (and its tilde version) are $\exp{i(I - 2\Pi)\phi}$ (and tilde version) projection-based rotations. Here $phi^{j}_k$ refers to the $k$-th component in the $j$-th phase list. Where $C$ is used, it refers to the circuit within the dashed box (or its conjugate transpose).} 
    \label{fig:qsvt_composition}
\end{figure}

Note for deep nesting that if one chooses to look at the block-encoding with respect to a different set of projectors (of which there is now no longer a canonical choice), then interpretation of the induced transformation changes. Again in our condensed superoperator notation, we will always refer to deep nesting as the map between the following pair of QSVT circuit descriptions
    \begin{equation}
        (\Phi_0, \tilde{\Pi}_0, \Pi_0, \ast), (\Phi_0, \tilde{\Pi}_1, \Pi_1, \ast),
    \end{equation}
and the following QSVT circuit description
    \begin{equation} \label{eq:deep_embedding}
        (\Phi_1, \tilde{\Pi}_1, \Pi_1, (\Phi_0, \tilde{\Pi}_0, \Pi_0,\ast)^\dagger\,\ast\!\ast),
    \end{equation}
where the two anonymous slots $(\ast)$ and $(\ast\ast)$ accept as inputs the first and second unknown unitary operations, as defined in Fig.~\ref{fig:qsvt_composition}. It is worth noting here that we have two input slots in the superoperator given in Eq.~\ref{eq:deep_embedding}, as opposed to the case of flat embedding. We can always resolve this by considering the partially applied function which fills one of these slots with a fixed signal; moreover, this will help address the apparent causal ambiguity in the resulting functional product discussed later, where two different nesting orders in programmable space can lead to the same product in functional space.

What remains is to identify the constraints under which the two concepts of QSVT nesting above can be made to satisfy properties allowing the diagram in Fig.~\ref{fig:qsp_as_natural_transformation} to commute. For flat nesting, this property will be almost identical to the QSP case, while for deep nesting, the required properties will be stronger, and the induced effect under the functor $T$ in Fig.~\ref{fig:qsp_as_natural_transformation} distinct.

\begin{theorem}[Flat (semantic) embedding for QSVT] \label{thm:qsvt_flat_embedding}
    Let $\Phi_0 \in \mathbb{R}^{n}$ and $\Phi_1 \in \mathbb{R}^m$ be antisymmetric phase-lists, and have them define two QSVT protocols, with shared projectors $\tilde{\Pi}, \Pi$. Then the flat nesting of $\Phi_0$ into $\Phi_1$ is equivalent to the \emph{flat embedding} of $\Phi_0$ into $\Phi_1$, also denoted $(\Phi_1\circ\Phi_0, \tilde{\Pi}, \Pi, U)$, and achieves the following transformation of the block-encoded $A = \tilde{\Pi}U\Pi$:
        \begin{equation}
            (G \circ F)^{SV}(\tilde{\Pi}U\Pi),
        \end{equation}
    where $(G \circ F)^{SV}$ denotes the application of the composition of $G$ (the polynomial transform encoded by $\Phi_1$) with $F$ (the polynomial transform encoded by $\Phi_0$) to the singular values of the general finite-dimensional linear operator $A$. Moreover, the resulting unitary protocol is itself remains \emph{flat embeddable}. Flat embedding satisfies the diagram given in Fig.~\ref{fig:qsp_as_natural_transformation} according to the assignments in Table~\ref{tab:qsvt_flat_categories} (polynomial composition). Proof is given in Appendix~\ref{sec:embedded_qsp_qsvt}. $\square$
\end{theorem}

The statement of flat embedding can succinctly described in terms of the diagram depicting natural transformations in Fig.~\ref{fig:qsp_as_natural_transformation}. Namely, the same antisymmetric symmetry imposed on the phases of QSP protocols which were semantically embeddable is imposed here through the lifting argument from QSP to QSVT. Consequently, for functional composition acting on the singular values of the block encoded operator $A$, the condition in the programmable space for the diagram to commute is near identical to that of standard QSP. For the same reason, the inner protocol remains contiguous, and therefore the resulting circuit truly treats QSVT subroutines as black boxes. The modified table for this assignment, in analogy to Table~\ref{tab:qsp_categories}, is given in Table~\ref{tab:qsvt_flat_categories}.

\begin{theorem}[Deep (semantic) embedding for QSVT with scalar block-encodings] \label{thm:qsvt_deep_embedding}
    Let $\Phi_0 \in \mathbb{R}^{n}$ and $\Phi_1 \in \mathbb{R}^m$ define two QSVT protocols, where $\Pi_0, \tilde{\Pi}_0$ are projectors encoding the scalar $F(|A_0|) = \tilde{\Pi}_0 U_0 \Pi_0$ for the block-encoding unitary of the inner protocol. Here $A_0$ denotes the image of $\tilde{\Pi}_0$, and $|A_0|$ its dimension. Then if $\Pi_1, \tilde{\Pi}_1$ also block encode a scalar, the deep nesting of $\Phi_0$ into $\Phi_1$ is equivalent to the \emph{deep embedding} of the two protocols, also denoted $(\Phi_1\land\Phi_0)$, and modifies the scalar block encoding in the following way:
        \begin{equation}
            \tilde{\Pi}_0 U_0 \Pi_0 = \sqrt{F(|A_0|)}
            \;\;\rightarrow\;\;
            U_1 \tilde{\Pi}_0 U_1^\dagger U_0\Pi_0 = \sqrt{|F(A_0)\,G(A_1)|},
        \end{equation}
    where $F, G$ are the polynomial transforms achieved by the protocols with phases $\Phi_0, \Phi_1$ respectively. Note that on the left-hand-side the block encoding is defined with respect to projectors $\Pi_0, \tilde{\Pi_0}$, while on the right-hand-side it is defined with respect to the modified projector $\tilde{\Pi}_0 \mapsto U_1\tilde{\Pi}U_1^\dagger$. If $F$ and $G$ are indicator functions which take value $1$ above $1/\sqrt{|A_0|}$ and $1/\sqrt{|A_1|}$ respectively, then the deep embedding block encodes $|A_0 \cap A_1|$, the (normalized) size of the intersection of the marked subsets of the two oracles. Moreover, the resulting unitary protocol itself remains \emph{deep embeddable}. Deep embedding satisfies the diagram given in Fig.~\ref{fig:qsp_as_natural_transformation} according to the assignments in Table~\ref{tab:qsvt_deep_categories} (set intersection or functional product). Proof is given in Appendix~\ref{sec:embedded_qsp_qsvt}. $\square$
\end{theorem}

For deep embedding, we take advantage of the fact that all of $\Pi_0, \Pi_1, \tilde{\Pi}_0, \tilde{\Pi}_1$ can be viewed as bifurcations on the relevant Hilbert space, inducing scalar block encodings. In this case their images and complements define two subspaces (marked and unmarked respectively), and the dimension of these subspaces (i.e., the images of $\Pi_0, \Pi_1$ for the marked subspaces) are therefore well-defined, notated $|A_0|, |A_1|$ respectively. As stated, the dimension of the intersection of these subspaces (the dimension of the image of the product of the marking projectors) can also be defined, and elements in these intersections prepared by the deeply nested protocol if $F, G$ are close to indicator functions. It is worth noting that in this case the action of the second functor $T$ in Fig.~\ref{fig:qsp_as_natural_transformation} is entirely different, as instead of polynomial composition, we take products, and that these products can often be interpreted as set operations. This is examined explicitly in the following section discussing distributed Grover search. In essence, by considering a different nesting operation and different restrictions on the projectors, we have found a distinct but algorithmically useful satisfying assignment for Fig.~\ref{fig:qsp_as_natural_transformation}. Note also that, unlike flat embedding, we know have two open arguments for our superoperator (the two scalars corresponding to the sizes of the marked subsets); if we bake-in the marked subset of the outer protocol, however, as we are always allowed to do, in turn partially applying the induced function, the seemingly causal action of the circuit (the outer protocol takes a subroutine, runs it where desired, and outputs a result) is restored.

\begin{table}[htpb]
    \centering
    \begin{tabular}{l l l}
        Category picture\;\;\; & QSVT picture & Description\\[-2.3ex]
        & &
        \\\hline
        \rule{0pt}{0.9\normalbaselineskip}$a, b, c$ & $\Phi, \tilde{\Pi}, \Pi$ & \text{QSVT phases and projectors}\\
        $f, g, h$ & $\Phi_1 \circ \Phi_0, \tilde{\Pi}, \Pi$ & \text{QSVT phase nesting} \\
        $Sa, Sb, Sc$ & $(\Phi, \tilde{\Pi}, \Pi, \ast)$ & \text{QSVT tuple (see Sec.~\ref{sec:condensed_notation_qsp_qsvt})}\\
        $Sf, Sg, Sh$ & $(\Phi_1 \circ \Phi_0, \tilde{\Pi}, \Pi, \ast)$ & \text{QSVT flat embedding (Eq.~\ref{eq:flat_embedding})}\\
        $Ta, Tb, Tc$ & $F^{SV}(A)$ & \text{Polynomial transform on s.v.'s}\\
        $Tf, Tg, Th$ & $F^{SV}(A) \rightarrow (G\circ F)^{SV}(A)$\;\;\;\; & \text{Polynomial composition on s.v.'s}\\
        $\tau a, \tau b, \tau c$ & $(\Phi, \tilde{\Pi}, \Pi, \ast) \rightarrow F^{SV}(A)$ & \text{Projection to block-encoded op}\\
    \end{tabular}
    \caption{Flat embedding for QSVT. As flat embedding (Thm.~\ref{thm:qsvt_flat_embedding}) is essentially the lifted version of semantic embedding for QSP, assignments to the diagram in Fig.~\ref{fig:qsp_as_natural_transformation} are similar, save with additional requirements on the orthogonal projectors, and how the composed functions are applied.}
    \label{tab:qsvt_flat_categories}
\end{table}

\begin{table}[htpb]
    \centering
    \begin{tabular}{l l l}
        Category picture\;\;\; & QSVT picture & Description\\[-2.3ex]
        & &
        \\\hline
        \rule{0pt}{0.9\normalbaselineskip}$a, b, c$ & $\Phi, \tilde{\Pi}, \Pi$ & \text{QSVT phases and projectors}\\
        $f, g, h$ & $\Phi_1 \land \Phi_0, \tilde{\Pi}^\prime, \Pi$ & \text{QSVT phase nesting} \\
        $Sa, Sb, Sc$ & $(\Phi, \tilde{\Pi}, \Pi, \ast)$ & \text{QSVT tuple (see Sec.~\ref{sec:condensed_notation_qsp_qsvt})}\\
        $Sf, Sg, Sh$ & $(\Phi_1, \tilde{\Pi}_1, \Pi_1, (\Phi_0, \tilde{\Pi}_0, \Pi_0,\ast)^\dagger\,\ast\!\ast)$\;\;\; & \text{QSVT deep embedding (Eq.~\ref{eq:deep_embedding})}\\
        $Ta, Tb, Tc$ & $F(|A_0|)$ & \text{Polynomial transformation on embedded scalar}\\
        $Tf, Tg, Th$ & $F(|A_0|) \rightarrow F(|A_0|)G(|A_1|)$ & \text{Polynomial multiplication on embedded scalars}\\
        $\tau a, \tau b, \tau c$ & $(\Phi, \tilde{\Pi}, \Pi, \ast) \rightarrow F(|A_0|)$ & \text{Projection to block-encoded scalar}\\
    \end{tabular}
    \caption{Deep embedding for QSVT. Note that for deep embedding (Thm.~\ref{thm:qsvt_deep_embedding}), there are now two relevant sets of projectors, with the inner protocol conjugating the left projector of the outer protocol. We also use different notation for abstract composition of QSVT phase lists in deep embedding, $(\Phi_1 \land \Phi_0)$, and note that this induces a different algebraic manipulation (product) of their respective polynomial transformations on embedded scalars.}
    \label{tab:qsvt_deep_categories}
\end{table}

While flat and deep embedding of QSVT exhibit different character in their underlying nesting procedure, we see that nevertheless both induce a simple operation on their underlying functional transforms. These induces operations (functional composition and functional products/set intersection) are simply describable according to a natural transformation between two functors (one describing circuit manipulation and the other functional manipulation). Consequently the category-theoretic abstraction introduced at the beginning of this work (Fig.~\ref{fig:qsp_as_natural_transformation}) can be seen to encompass far larger algorithmic aims than just the composition of polynomials embedded in QSP protocols. Expanding the settings in which the requirements for this diagram to commute are satisfied is a major avenue for extension of this work, and opens the possibility for sophisticated, purely functional interpretations of a wide class of quantum computations and circuit ansätze.

Before moving on to a description of applications of semantic embedding in QSVT to known quantum algorithms, it is worthwhile to review the question of whether the functional manipulations discussed here constitute a `complete set' of manipulations. Toward this end, we note that the polynomial transforms achieved by both QSP and QSVT are (considering only the real part of $P$ for a moment, which is almost freely chooseable) those with (1) definite parity in $x$ and (2) 1-bounded norm on the interval $x \in [-1,1]$. These constraints are due to fundamental symmetries of the QSP ansatz, as well as the unitarity of the overall evolution, with derivations of these properties implicit in \cite{gslw_19}. A useful way to characterize the set of possible manipulations is thus to determine which manipulations of two functions satisfying this property produce a third function also satisfying this property. Writing down such a conjectural list of such manipulations is simple enough:
\begin{align}
    (F, G, x) &\mapsto (G \circ F)(x),\label{eq:manipulation_1}\\
    (F, G, x_0, x_1) &\mapsto F(x_0)G(x_1),\label{eq:manipulation_2}\\
    (F, G, x) &\mapsto \alpha F(x) + \beta G(x), \quad |\alpha| + |\beta| \leq 1.\label{eq:manipulation_3}
\end{align}
In each of these cases, the resulting function (when applied to the signal(s)) necessarily has definite parity and bounded norm. In the second case, where there are two possible arguments, fixing either one to be a constant results in the desired properties, and consequently we can say that in that case as well we remain within the permitted space of QSP/QSVT-achievable functions. Evidently flat and deep QSVT embedding correspond to the first two manipulations, while the third, due to its simpler, linear nature in the two functions, can be achieved via the simple combination of QSVT subroutines using linear combinations of unitary (LCU) methods \cite{cw_lcu_12} with constant additional space (it is worthwhile to note that this protocol cannot be a nested or intrinsically ordered one, as both functions are applied to the same argument).

To what extent do these constitute a generating set for all possible norm and parity preserving combinations of two functions (and their respective argument lists)? Evidently we have identified a desired set of (possibly multivariable) polynomials whose restriction to any one free variable satisfies two desired properties (parity and norm on an interval). It turns out that this question is almost trivial given how we have posed the problem: the underlying operations available to us in the monoid of polynomials are precisely composition (\ref{eq:manipulation_1}) and multiplication (\ref{eq:manipulation_2}), while viewing the polynomials as a module, also permit linear combinations according to properly normalized scalars (\ref{eq:manipulation_3}). If we are given an arbitrary `manipulation rule' as above, \emph{along with the promise that said result is `constituted' from the two argument functions}, then this `constitution' procedure must necessarily arise from manipulations (i.e., binary operations) permitted in the underlying monoid/vector space. While it may be possible to assign additional structure to the relevant space of polynomials, for the purpose of most quantum computational problems, the monoids agreeing with (\ref{eq:manipulation_1}-\ref{eq:manipulation_2}) and the module agreeing with (\ref{eq:manipulation_3}), appear to exhaust the privileged structure of the underlying space of functions.

\section{Applications in known quantum algorithms} \label{sec:applications}

We identify previous cases where quantum circuits are self-embedded to solve concrete problems, and recast them as instances of semantic embedding. To showcase differing interpretations of this recasting, we focus on three instances: (1) distributed scheduling \cite{grover_scheduling_02}, (2) communication complexity separations for linear algebra problems \cite{ms_communication_complexity_22}, and (3) the soundness of certain succinct argument protocols against quantum adversaries \cite{cmsz_qpsa_2022, lombardi_pqzk_2021}.

\subsection{Quantum scheduling and oblivious amplitude amplification}
    
    Unstructured search and its generalization, amplitude amplification (AA), are well-studied quantum algorithmic subroutines \cite{grover_05, nc_01}. In many ways, AA looks like a restricted instance of QSP/QSVT: two unitary operations are interleaved and produce, at a given circuit depth, a desired transition with high probability. For Grover search the interleaved operators depend on the following projectors:
        \begin{equation}
            \tilde{\Pi} = H^{\otimes n} |0\rangle\langle 0 | H^{\otimes n},
            \quad
            \Pi = |m\rangle\langle m|,
        \end{equation}
    where $|m\rangle$ is the equal superposition over \emph{marked states} in the computational basis, where this marking is done by the standard Grover oracle $|k\rangle \mapsto (-1)^{\delta_{mk}}|k\rangle$. Grover's algorithm is the application of the following product of unitaries to the state $H^{\otimes n}|0\rangle$,
        \begin{equation} \label{eq:interleaved_grover}
            W 
            = 
            \prod_{k} e^{i\pi (I - 2\tilde{\Pi})}e^{i\pi (I - 2\Pi)},
        \end{equation}
    where $k$ ranges over some set of size $\mathcal{O}(\sqrt{n})$. In other words one alternately reflects around the uniform superposition (the initial state) and the marked subspace. These reflections generate a rotation toward the marked subspace, and the size of this rotation determines the runtime. The required number of such alternating reflections is quadratically fewer than might be assumed; in general this improvement is known to be optimal for otherwise unstructured data, and strongly believed to exist for realistic problems such as CircuitSAT.
    
    We briefly give definitions and theorems for fixed-point amplitude amplification and oblivious amplitude amplification, so that the discussion of their variants for multiple marking oracles later will make sense. These theorems are quite pared down, following a long line of simplifying work for amplitude amplification \cite{hoyer_00, grover_05, ylc_14, lyc_16_equiangular_gates, gslw_19}. In fixed-point amplitude amplification (Thm.~\ref{thm:fixed_point_aa}), the computing party alternates reflections about a known initial state and a marked subspace, while in oblivious amplitude amplification (Thm.~\ref{thm:oblivious_aa}), the known initial state is replaced by a state prepared by a (possibly oracular) isometry.
    
    \begin{theorem}[Fixed point amplitude amplification] \label{thm:fixed_point_aa}
        Theorem 27 in \cite{gslw_19}. Let $U$ be a unitary and $\Pi$ an orthogonal projector such that $a|\psi_G\rangle = \Pi U |\psi_0\rangle$ and $a > \delta > 0$. Here we mean that $U$ acting on $|\psi_0\rangle$ produces some small but non-zero overlap $a$ with the good state $|\psi_G\rangle$. Then there is a unitary circuit $\tilde{U}$ such that $\lVert |\psi_G\rangle - \tilde{U}|\psi_0\rangle\rVert \leq \epsilon$, which uses a single auxiliary qubit and $\mathcal{O}(\delta^{-1}\log{\epsilon^{-1}})$ uses of $U$, $U^\dagger$, $C_{\Pi}NOT$, $C_{|\psi_0\rangle\langle\psi_0|}NOT$ and $e^{i\phi\sigma_z}$ gates. Proof follows by noting that
            \begin{equation}
                \tilde{\Pi}U\Pi = a|\psi_G\rangle\langle\psi_0|
            \end{equation}
        is a scalar block encoding of the overlap $a$, where $\tilde{\Pi} = |\psi_G\rangle\langle \psi_G|$, and choosing a minimal degree polynomial which is $\epsilon$-close to $1$ at arguments above $\delta$. $\square$
    \end{theorem}
    
    \begin{theorem}[Oblivious amplitude amplification] \label{thm:oblivious_aa}
        Modified from Theorem 28 of \cite{gslw_19}. Let $U$ be a unitary, $\epsilon > 0$, $a > \delta > 0$, $\tilde{\Pi}$, $\Pi$ orthogonal projectors, and $W: \text{Img}(\Pi) \mapsto \text{Img}(\tilde{\Pi})$ an isometry such that
            \begin{equation}
                \lVert a W|\psi\rangle - \tilde{\Pi} U |\psi\rangle\rVert \leq \epsilon
            \end{equation}
        for all $|\psi\rangle \in \text{Img}(\Pi)$. Then we can construct a unitary $\tilde{U}$ such that for all $|\psi\rangle \in \text{Img}(\Pi)$
            \begin{equation}
                \lVert W|\psi\rangle - \tilde{\Pi} \tilde{U} |\psi\rangle\rVert \leq \epsilon.
            \end{equation}
        The circuit $\tilde{U}$ uses $\mathcal{O}(\delta^{-1}\log{\epsilon^{-1}})$ uses of $U$, $U^\dagger$, $C_{\Pi}NOT$, $C_{\tilde{\Pi}}NOT$, and single qubit rotation gates. Proof is almost identical to the non-oblivious setting (Thm.~\ref{thm:fixed_point_aa}), where $aW = \tilde{\Pi}U\Pi$ is the block encoded operator, and the desired polynomial sends all arguments above $\delta$ to within $\epsilon$ of $1$. $\square$
    \end{theorem}
    
    Having established the basic map between QSVT and standard Grover search (through AA), we now discuss a distributed variant of the same process. In Grover's scheduling paper \cite{grover_scheduling_02} he considers a setting in which two parties each have their own marking oracle, privileging two sets of indices $m, m^\prime$. Their goal is, with minimal resources, to prepare a state in the intersection $m \cap m^\prime$ (here overloading the variables $m, m^\prime$). To do this Grover constructs a nested protocol consisting of an \emph{inner protocol}, call it $U_m$ and an \emph{outer protocol}, call it $U_{m^\prime}$ which uses $U_m$ as a subroutine, defined as follows:
        \begin{align}
            U_m &= 
            \prod_{k}e^{i\pi (I - 2\tilde{\Pi})}e^{i\pi (I - 2\Pi_{m})},\\
            U_{m^\prime} &=
            \prod_{k}
            \left[U_{m}e^{i\pi (I - 2\tilde{\Pi})}U_{m}^\dagger\right]
            e^{i\pi (I - 2\Pi_{m^{\prime}})},
        \end{align}
    where the entire protocol, summarized in the outer protocol, is applied to the state $U_{m}H^{\otimes n}|0\rangle$. Here $\tilde{\Pi}$ is as before the projector onto the uniform superposition on $\log{n}$ qubits for both parties, while $\Pi_{m}, \Pi_{m^\prime}$ are projectors for the marked subspaces.
        
    For the inner protocol this is just AA using QSVT with a block-encoded scalar $A = a = \sqrt{m/n}$
        \begin{equation} \label{eq:inner_protocol}
            U_{m} \equiv
            \begin{bmatrix}
                \;A & \cdot\;\; \\ 
                \;\;\cdot & \cdot\;\; 
            \end{bmatrix}.
        \end{equation}
    On the other hand, the outer protocol represents a more sophisticated process. In this case, the block-encoded operator is now \emph{dependent} on that of the inner protocol, namely
        \begin{equation} \label{eq:outer_protocol}
            U_{m^\prime} \equiv
            \begin{bmatrix}
                \;A^\prime & \cdot\;\; \\ 
                \;\;\cdot & \cdot\;\; 
            \end{bmatrix}.
        \end{equation}
    where $A^\prime = a^\prime$ is a scalar whose value is $\sqrt{(m\cap m^\prime)/n}$, clearly dependent on the size of the intersection of the marked sets (as is to be expected for a scheduling problem interested in probing and preparing elements in this intersection). In other words, we see that \emph{unitary conjugation of another party's projectors}, as used in the projection-controlled-NOTs ubiquitous in QSP, can have the effect of \emph{computing joint functions on block-encoded data} with respect to two (or more) local oracles. This is precisely an instance of deep embedding of QSVT protocols as given in Thm.~\ref{thm:qsvt_deep_embedding}. Moreover, as is discussed in \cite{grover_scheduling_02} the round and communication complexities of this protocol are easily computed.

\subsection{Communication complexity for distributed quantum algorithms}
        
    We now place semantic embedding in the context of a recent result \cite{ms_communication_complexity_22} concerning communication complexity separations between quantum and classical algorithms for linear algebraic problems. This in turn bolsters the following idea: distributed quantum computation problems where the parties can make use of quantum channels for communication are a \emph{natural setting} in which to understand the utility of semantic embedding.
    
    Communication complexity is a well studied field in both classical and quantum computer science. Take, for instance, a setting where two separated parties, one holding some data $x$ and another holding some data $y$, seek to compute some joint function of their data $f(x, y)$, and moreover wish to do this using as little communication as possible. Communication complexity often refers to the minimal required communication (usually in bits, or qubits) to perform a particular algorithmic task (say, to compute $f(x, y)$ to some specified accuracy, with some specified probability). Studying this complexity is reasonable when the cost of the computing done by each party is far cheaper relative to the cost of communication between parties.
    
    While the detailed work of \cite{ms_communication_complexity_22} discusses multi-party protocols and a diverse set of communication models, we restrict our discussion to one and two-way communication protocols between two parties attempting to produce an approximation to the solution to a set of linear equations with high probability. We reproduce a simplified version of their problem statement below, before describing their solution in terms of self-embedded QSVT, and then giving comments on extensions to their setting.
    
    \begin{problem}[Distributed matrix inversion] \label{problem:communication_complexity}
        Let two parties with quantum computers, Alice and Bob, be such that Alice holds some $A \in \mathbb{R}^{m\times n}$ and Bob holds some $b \in \mathbb{R}^m$. The goal of the parties is to output the state $|A^{+}b\rangle$ using as little communication (in qubits) as possible. Additionally, we will often refer to the condition number (minimal singular value) of $A$ by $\kappa$, and the cosine of the angle between $b$ and the column space of $A$ by $\gamma \equiv \lVert AA^{+}b \rVert/\lVert b\rVert$. For the classical analogue of the problem, the goal is to output samples from the proper state. Here $A^{+}$ is the pseudoinverse of the matrix $A$. $\square$
    \end{problem}
    
    \begin{theorem}[Theorems 4 and 9 from \cite{ms_communication_complexity_22} together, compressed] \label{thm:communication_complexity}
        Take two parties holding $A$ and $b$ as given in Problem~\ref{problem:communication_complexity}. Then the quantum communication complexity of outputting $|A^{+}b\rangle$ is the following, according to the communication model
            \begin{enumerate}
                \item One-way communication (Alice $\mapsto$ Bob). Bounded above by $\mathcal{O}(\log{[mn]}\min{[m, n]}\,\kappa^2/\gamma^2)$.
                \item One-way communication (Bob $\mapsto$ Alice). Bounded above by $\mathcal{O}(\log{[m]}\,\kappa^2/\gamma^2)$.
                \item Two-way communication. Bounded above by $\mathcal{O}(\log{[m]}\,\kappa/\gamma)$.
            \end{enumerate}
        Moreover, for the task of two classical parties attempting to sample from $A^{+}b$, the following lower bounds are known
            \begin{enumerate}
                \item One-way communication (Alice $\mapsto$ Bob). Bounded below by $\Omega(\min{[m, n]}\log{[\min{[m, n]}]})$.
                \item One-way communication (Bob $\mapsto$ Alice). Bounded below by $\Omega(\min{[m, n]})$.
                \item Two-way communication. Bounded below by $\Omega(\min{[m, n]})$.
            \end{enumerate}
        Proofs of these complexities are contained in the referenced work. $\square$
    \end{theorem}
    
    We briefly describe the quantum protocols which achieve the communication complexity provided in the statement of Thm.~\ref{thm:communication_complexity}. Given that Alice holds a description of a matrix and Bob holds a description of the quantum state, the simplest setting is that of one way communication from Bob to Alice. In this case there is only one round of communication wherein Bob sends $|b\rangle$ (a total of $\log{[m]}$ qubits) to Alice, who applies the QSVT protocol which block-encodes the pseudoinverse $A^{+}$ to the state $|b\rangle$ (following Theorem 41 of \cite{gslw_19}) and measures. The probability of obtaining $|A^{+}b\rangle$ is $\gamma^2/\kappa^2$, and thus repetition by the inverse of this probability yields the desired bound. For one-way communication from Alice to Bob, Alice makes use of the Choi-Jamiołkowski isomorphism to send a state to Bob encoding the matrix $A$, which requires a rescaling by the Frobenius norm, in addition to which post-selection requires the stated $\mathcal{O}(\log{[mn]}\min{[m, n]}\,\kappa^2/\gamma^2)$ communication complexity. Finally, for the two-way communication case, the parties use oblivious amplitude amplification according to the reflection about the target state
        \begin{equation}
            2|\psi\rangle\langle\psi| - I = U\biggr[2|0\rangle|b\rangle\langle b|\langle0| - I\biggr]U^\dagger,
        \end{equation}
    where $U$ is the unitary which block-encodes $A^{+}$, and additional registers beyond $|b\rangle$ have been initialized to $|0\rangle$. Consequently, by simple application of Grover amplification, this technique permits a quadratic improvement in terms of the original success probability $\gamma^2/\kappa^2$ as indicated in Theorem~\ref{thm:communication_complexity}.
    
    It is now relatively clear how to interpret the protocols in \cite{ms_communication_complexity_22} (in the two-way communication setting) in terms of semantic embedding. Important to note, however, is that the action applied by Bob in this setting is just a reflection about his supplied state, summarized by $2|b\rangle|0\rangle\langle b|\langle 0| - I$. This is, in our language, a trivial QSVT protocol of length one, with phase list $\Phi_0 = \{0, 0\}$, where the unitary applied by Bob simply block-encodes itself. Consequently the resulting deep embedding (Thm.~\ref{thm:qsvt_deep_embedding}), defined according to the functional map
        \begin{equation}
            (\Phi_1, \tilde{\Pi}_1, \Pi_1, (\Phi_0, \tilde{\Pi}_0, \Pi_0, \ast)^\dagger\,\ast\!\ast),
        \end{equation}
    has $\Phi_0$ effectively trivial, and $\Phi_1$ the same set of phases as for fixed-point amplitude amplification. Here the projectors $\tilde{\Pi}_1, \Pi_1$ are onto the target state $|b\rangle$ and the uniform superposition respectively, while $\tilde{\Pi}_1, \Pi_1$ are onto the state $|0\rangle$ and the uniform superposition respectively. Additionally, the argument unitary taken by $(\ast\ast)$ is the unitary $U$ which block-encodes $A^{+}$, while the argument unitary taken by $(\ast)$ is the identity. While this is not a particularly sophisticated use of semantic embedding, the flexibility of its application suggests a variety of related questions in the study of communication complexity.
    
    An intriguing prospect is whether there exist substantively more complex interactive protocols in the model considered in \cite{ms_communication_complexity_22}, namely those which rely on QSVT phase lists other than that for amplitude amplification. In the language of QSVT, amplitude amplification corresponds to an embedded polynomial function which is approximately constant over nearly all arguments. There is no reason, however, that interactive protocols be restricted to computing such functions (which are approximately the product of two single-variable functions) \cite{mgb_coherent_arithmetic_22, rossi_m_qsp_22}. QSVT seems to offer the only current method for well-approximating such `non-factorable' transforms, though as stated for amplitude amplification many sufficiently efficient alternative techniques already exist.
    
\subsection{Quantum succinct arguments}
    
    While quantum computing is perhaps better known for posing a threat to common cryptographic constructions, quantum algorithmic techniques have also had success in investigating cryptographic constructions with supposed security against even quantum adversaries. This section gives a light introduction to settings in quantum cryptography that make implicit use of semantic embedding to prove the quantum security of certain cryptographic constructions. There is also indication that such constructions are merely the first in a larger, imminent class of results.
    
    Recent work \cite{cmsz_qpsa_2022} considers the post-quantum security of a succinct argument system in the standard model when instantiated with various classical cryptographic objects known to exist under the assumed post-quantum hardness of learning with errors (LWE) \cite{regev_09}. A key topic of investigation in this work is the quantum analogue of a common classical method for proving the security of cryptographic schemes: rewinding.
    
    The work of \cite{cmsz_qpsa_2022} considers a specific succinct argument system, Killian's protocol \cite{kilian_92}, by which a collision-resistant hash function is used to transform a probabilistically checkable proof (PCP) into an interactive protocol with exponential savings in communication complexity, compared to simply querying the PCP. This improvement comes at the cost of a computational assumption for the soundness (i.e., the ability for the proving party to fool the verifier is reduced to the assumed difficulty of a computational problem). The classical security of this protocol is proven via a common technique, \emph{rewinding}, whereby a verifer's or prover's state is saved midway through a hypothetical protocol and rerun until many accepting transcripts are collected (and from which the long PCP string is extracted). The genericness of this definition follows from that rewinding is a somewhat vague term, and can refer to multiple independent settings; a common feature is oracle access to the actions of one party in an interactive protocol. For reasons specific to quantum mechanics, these methods fail when attempting to prove security in the quantum setting. Casually, this is because of the inability to clone arbitrary states, as well as the often destructive nature of quantum measurements (forbidding a quantum party from rerunning an adversary from a consistent saved state).
    
    The major contribution of \cite{cmsz_qpsa_2022} is a construction showing, under the assumed existence of so-called collapsing hash functions \cite{unruh_collapsing_2016}, that Killian's protocol is post-quantum secure. The key subroutines relating to QSVT in the proofs of security in \cite{cmsz_qpsa_2022} are \emph{state recovery} and \emph{state repair}, which together circumvent the apparent impossibility of rewinding a quantum adversary. We informally summarize these techniques below, and discuss their connection to semantic embedding.
    
    \begin{definition}[State recovery in \cite{cmsz_qpsa_2022} (Informal)] \label{def:state_recovery}
        A key desire in quantum rewinding protocols is the ability to recover a state which has been disturbed by an intervening destructive measurement. The paper first posits the ability to perform a projective measurement $\text{Equals}_{|\psi\rangle}$, for $|\psi\rangle$ the intermediate state of the prover, and shows that alternating this projective measurement with some $B$, an intervening projective measurement querying the desired information for the transcript, can be used to recover $|\psi\rangle$ with high probability quickly through consequences of Jordan's lemma (Lemma~\ref{lemma:joran}). Unfortunately $\text{Equals}_{|\psi\rangle}$ is impossible (by no-cloning) to efficiently implement. Their solution is to thus to relax \emph{state recovery} to \emph{state repair} (Def.~\ref{def:state_repair}) in which only \emph{those aspects of the state necessary for the relevant proof of security} are corrected. $\square$
    \end{definition}
    
    \begin{lemma}[Jordan's lemma \cite{jordan_75}] \label{lemma:joran}
        Variously discussed in \cite{cmsz_qpsa_2022, mw_05, zhandry_20, lombardi_pqzk_2021, gslw_19, regev_06, cs_qsvt_tang_tian}. For any two Hermitian projectors $\Pi_A, \Pi_B$ on a Hilbert space $\mathcal{H}$, there exists an orthogonal decomposition for this space $\mathcal{H} = \bigoplus_{k} S_j$ into one- and two-dimensional subspaces, $S_j$ (the Jordan subspaces), where each is invariant under both $\Pi_A$ and $\Pi_B$. Moreover: (1) in each one-dimensional subspace $\Pi_A$ and $\Pi_B$ act as the identity or a rank-zero projector and (2) in each two-dimensional subspace, $\Pi_A$ and $\Pi_B$ are rank-one projectors, and there exist distinct orthogonal bases for each $S_j$, here denoted $\{|v_{j, 0}\rangle, |v_{j_1}\rangle\}$ and $\{|w_{j, 0}\rangle, |w_{j_1}\rangle\}$ such that $\Pi_A$ and $\Pi_B$ project onto $|v_{j, 0}\rangle$ and $|w_{j, 0}\rangle$ respectively. These are the left and right singular vectors of $\Pi_A \Pi_B$ respectively, with singular value $s_j = |\langle v_{j, 0} | w_{j,0}\rangle|$. A proof of this lemma can be found in \cite{regev_06}, and more accessibly in \cite{cs_qsvt_tang_tian}. $\square$
    \end{lemma}
    
    \begin{definition}[State repair in \cite{cmsz_qpsa_2022} (Informal)] \label{def:state_repair}
        While the implementation of $\text{Equals}_{|\psi\rangle}$ in Def.~\ref{def:state_recovery} is not efficient, the requirement asked of the rewinding procedure in \cite{cmsz_qpsa_2022} is only that the success probability for extraction does not decay with repeated rewinding attempts. Thus \cite{cmsz_qpsa_2022} defines a new projective measurement $\text{Test}_{\epsilon}$, whose image is the subspace whose elements which have a success probability at least $\epsilon$. Repeatedly interleaving this test with the randomized measurement $A_r$ eventually, upon $\text{Test}_{\epsilon}$ returning $1$, restores the state's success probability. 
        
        Unfortunately $\text{Test}_{\epsilon}$ is not efficiently implementable either, leading the work to introduce $\text{ApproxTest}_{\epsilon,t}$. This itself is implemented by a series of alternating projective measurements (as well as a post-selective aspect, as it can be shown that $\text{ApproxTest}_{\epsilon,t}$ is not by itself perfectly projective). The parameter $t$ here represents a number of trials which are used to estimate whether the success probability on which the $\text{Test}_{\epsilon}$ projective measurement thresholds is being exceeded or not. It is shown that, even relaxing projective to almost projective measurements, and relaxing $\text{Test}_{\epsilon}$ to $\text{ApproxTest}_{\epsilon,t}$ (the former itself relaxing what was desired from $\text{Equals}_{|\psi\rangle}$), the required properties of the rewinding protocol are preserved. $\square$
    \end{definition}
    
    To summarize \cite{cmsz_qpsa_2022}, we define the projectors given in their state repair and recovery procedures, and highlight the deep embedding of QSVT. At a skeletal level the paper considers two alternating projector sequences. The second interleaved sequence (approximately projecting onto the \emph{good-success-probability} subspace) is used as a subroutine by the first (which with high probability returns a state to that good subspace after extracting a valid transcript).
    
    \begin{align}
        (U_1 \land U_0) &\equiv \text{ApproxTest}_{\epsilon, t}, A_{r}, \text{ApproxTest}_{\epsilon, t}, A_{r}, \cdots \\
        U_0 &\equiv \text{CProj}, M_{|+R\rangle}, \text{CProj}, M_{|+R\rangle}, \cdots
    \end{align}
    
    Here the originally desired $\text{Equals}_{|\psi\rangle}$, which would have tested for the desired mid-protocol state of the adversary has been replaced by $\text{Test}_{\epsilon}$, which only projects onto states with high enough success probability in the task given to the adversary. This has in turn been replaced by $\text{ApproxTest}_{\epsilon, t}$, which approximates the behavior of $\text{Test}_{\epsilon}$. Here $A_r$ is the challenge to the adversary on randomness $r$ through whose measurement one is perhaps both obtaining an accepting transcript but also damaging the adversary's state. As given, to generate $\text{ApproxTest}_{\epsilon, t}$ a separate amplification procedure is necessary, alternating $\text{CProj}$ and $M_{|+R\rangle}$. Here $\text{CProj}$ (a controlled-projector) measures $\{\Pi_{r}, I - \Pi_r\}$ coherently depending on $r$, while $M_{|+R\rangle}$ measures according to the projectors
        \begin{equation}
            \{|+R\rangle\langle +R|,\, I - |+R\rangle\langle +R|\},
        \end{equation}
    where $|+R\rangle$ is the uniform superposition over the possible random $r$.
    
    \begin{remark}[Caveat on the differences between alternating projective measurements (i.e., Watrous technique \cite{mw_05}) and amplitude amplification \cite{gslw_19}]
        Going back to the original statements of unstructured search and amplitude amplification \cite{hoyer_00, grover_05}, the algorithm has been posed as either (1) a series of interleaved reflections about relevant subspaces (some oracularly provided), and (2) a series of projective measurements onto these subspaces, inducing a random walk. For our purposes, the runtimes of algorithms attempting to prepare oraculalry marked states in both of these ways are effectively the same, as remarked in the improvements to \cite{cmsz_qpsa_2022} given in \cite{lombardi_pqzk_2021}. Wherever the Watrous technique is applied, QSVT can be inserted almost seamlessly. $\square$
    \end{remark}
    
    To summarize, the nested protocol has allowed the rewinding party to generate many accepting transcripts on different inputs, using ApproxTest to suitably approximate Test, which itself projects onto a subspace with properties that are \emph{good enough} (i.e., high enough success probability over random challenges $r$) to avoid having to have used Equals. In turn, the way in which these protocols were nested was precisely the deep embedding construction given in Thm.~\ref{thm:qsvt_deep_embedding}. Even in this involved instance, semantic embedding allows one to identify that the key insight is simply being able to prepare a quantum state according to the intersection of two marking oracles (which were themselves based on only approximate projectors). We finally note that, unlike in the communication examples given previous, the division between inner and outer protocols in this application of semantic embedding is used to outline a hierarchy of computational tasks, rather than to accommodate a physical separation between computing parties.

\section{Discussion and conclusion}

In this work we have introduced the concept of \emph{semantic embedding} for quantum algorithms using QSP/QSVT. That is, observing that QSP and QSVT as quantum algorithms can be described in terms of the polynomial transforms they apply, we provide methods for manipulating and combining QSP/QSVT quantum circuit ansätze to induce analogous algebraic manipulations of their embedded polynomial transforms. In category theoretic terms, we identify a natural transformation between two functors (one depicting manipulations of quantum circuits, and the other depicting polynomial transformations), and determine properties of arrows between objects representing QSP/QSVT phase lists \emph{such that this diagram commutes}. Alternatively, we provide a series of simple conditions under which the pre-image of functional operations under this natural transformation is efficiently computable and respects the contiguity of subroutines.

Returning to the original motivation for this work, these are a set of non-trivial rules for the construction of highly expressive quantum circuits such that the algorithmist, who is concerned only with the efficient serial application of functions to meaningful data, can rest assured that their semantic intentions are naturally preserved and accessible within a representing quantum circuit.

For standard QSP, the property required on the category of phase lists is that they obey a special symmetry under reversal and negation. For QSVT, we show there are multiple choices of constraint (corresponding to flat and deep embedding) depending on the desired polynomial manipulation. We leave open the possibility of further functional manipulations using these circuit ansätze as basic units, but give argument that those presented are exhaustive up to reasonable definition (I.e., as per Eqs.~\ref{eq:manipulation_1}-\ref{eq:manipulation_3}). Parallel to this work, we advocate the general study of how choice of parameterized circuit ansätze relates to achievable functional transforms. Understanding simple (but still verifiably expressive) circuit ansatz holds great promise \cite{sym_qsp_21}, and represents a path by which QSP/QSVT can be made more helpful in the near term.

Finally, we discuss known quantum algorithms where semantic embedding already occurs (albeit implicitly). In two of these three cases (distributed search and distributed linear algebraic problems) two computing parties exist, separated in space, with one party submitting a quantum state repeatedly to the other to perform a quantum computational subroutine. In the third instance (proofs of security for succinct arguments) a party is afforded black-box access to a unitary operation, and embeds subroutines successively for semantic aims, breaking down a complex computation into a series of simpler steps. Consequently we give two regimes in which semantic embedding has utility: (1) when there physically exist two (or more) separated computing parties, each of whom possesses some computational responsibility, and (2) when there exists only one party, but the computing task is more easily analyzed when broken into hierarchical (or nested) stages. A key observation is that the pre-image of functional composition and products as given in this work \emph{preserve the separation of subroutines in circuit space} necessary to give well-founded circuits for (1) and (2).

Foremost we intend that the clarity of statements like that of semantic embedding, i.e., simple conditions under which manipulations in the \emph{programmable space} of quantum circuits automatically induce analogous manipulations in the \emph{functional space}, can foster more ambitious quantum algorithm design rooted in higher-order reasoning about functional transforms. Likewise, better understanding how constraints on circuit parameterizations interact with algorithmic expressivity, and developing more diverse families of QSP/QSVT-like parameterized ansätze \cite{sym_qsp_21, rossi_m_qsp_22, dlnw_infinite_qsp_22}, can provide modules directly compatible with this work. In turn, we offer that the celebrated unifying aspect of QSVT could be re-positioned as a branching point for the development of simple and practical \emph{families} of quantum algorithms bound by simple, category-theoretically described rules.

\vspace{-1em}

\section{Acknowledgements}

The authors would like to thank Bill Munro, Victor Bastidas, and John Martyn for helpful comments. ZMR was supported in part by the NSF EPiQC program. IC was supported in part by the U.S. DoE, Office of Science, National Quantum Information Science Research Centers, and Co-design Center for Quantum Advantage (C2QA) under contract number DE-SC0012704.

\bibliography{main}

\appendix

\section{Basics of QSP and QSVT} \label{sec:qsp_qsvt_theory}

In this section we give the common presentations of the main theorems of quantum signal processing (QSP) and quantum singular value transformation (QSVT), citing relevant major works in their development. These algorithms take as input repeatable unitary processes and produce as output unitary processes whose dependence on the input unitary process (and its possibly unknown underlying parameters) is both complex and precisely controllable. The notation used here is standard unless otherwise noted, though competing conventions are discussed in the appendix of \cite{mrtc_21} for those interested.

\subsection{The theory of QSP and lifting to QSVT} \label{sec:lifting_qsp_qsvt}

We give simplified statements of the major theorems of QSP, focusing on how they are often applied to useful algorithmic problems. We then sketch the lifting argument used to derive QSVT from QSP, which is referred to in the main text.

\begin{definition}[QSP protocol \cite{lyc_16_equiangular_gates, lc_17_simultation, lc_19_qubitization, gslw_19, haah_2019}] \label{def:qsp_protocol}
    Let $\Phi \in \mathbb{R}^{d + 1}$. A QSP protocol is a product of rotations in SU(2),
        \begin{equation}
            U_{\Phi} \equiv e^{i\phi_{0}\sigma_z}\prod_{k = 1}^{d} [e^{i\arccos{x}\sigma_{x}}e^{i\phi_{k}\sigma_z}],
        \end{equation}
    where $\phi_{k}$ is the $k$-th element of $\Phi$. Often $e^{i\arccos{x}\sigma_{x}}$ is denoted $W(x)$, where the \emph{signal} $x$ is such that $x \in [-1, 1]$. The unitary $W(X)$ is often assumed unknown but query-accessible, while the $\phi_k$ are chosen by the computing party. Note here and elsewhere $\sigma_x, \sigma_z$ are the usual single-qubit Pauli matrices with non-zero entries $\pm 1$, i.e.,
        \begin{equation} \label{eq:pauli_matrices}
            \sigma_x \equiv 
            \begin{bmatrix}
                \;0 & 1\;\\
                \;1 & 0\;
            \end{bmatrix},
            \quad\quad
            \sigma_z \equiv 
            \begin{bmatrix}
                \;1 & 0\;\\
                \;0 & -1\;
            \end{bmatrix},
        \end{equation}
    which generate rotations about orthogonal axes on the Bloch sphere. $\square$
\end{definition}

\begin{theorem}[Unitary form of quantum signal processing (QSP), i.e., $\Phi \mapsto P, Q$] \label{thm:forward_qsp}
    Theorem 3 in \cite{gslw_19}. Let $d \in \mathbb{N}$. There exists $\Phi = \{\phi_0, \phi_1, \cdots, \phi_d\} \in \mathbb{R}^{d + 1}$ such that for all $x \in [-1, 1]$:
        \begin{equation} \label{eq:qsp_def}
            U_{\Phi}(x) 
            = 
            e^{i\phi_0\sigma_z}\prod_{k = 1}^{d} \left[W(x)\, e^{i\phi_k \sigma_z}\right]
            =
            \begin{bmatrix}
                P(x) & i\sqrt{1 - x^2}Q(x)\\
                i\sqrt{1 - x^2}Q^*(x) & P^*(x)
            \end{bmatrix},
        \end{equation}
    if and only if $P, Q \in \mathbb{C}[x]$ such that
        \begin{enumerate}[label=(\arabic*)]
            \item $\text{deg}(P) \leq d$ and $\text{deg}(Q) \leq d - 1$.
            \item $P$ has parity-$d \pmod 2$ and $Q$ has parity-$(d-1)\pmod 2$.
            \item For all $x \in [-1, 1]$ the relation $|P(x)|^2 + (1 - x^2)|Q(x)|^2 = 1$ holds. $\square$
        \end{enumerate}
\end{theorem}

\begin{theorem}[Reconstruction of QSP protocols from partial specification, i.e., $\tilde{P}, \tilde{Q} \mapsto \Phi$] \label{thm:backward_qsp}
    Theorem 5 in \cite{gslw_19}. Let $d \in \mathbb{N}$ fixed. Let $\tilde{P}, \tilde{Q} \in \mathbb{R}[x]$. There exists some $P, Q \in \mathbb{C}[x]$ satisfying conditions (1-3) of Theorem~\ref{thm:forward_qsp} such that $\tilde{P} = \Re(P)$ and $\tilde{Q} = \Re(Q)$ if and only if $\tilde{P}, \tilde{Q}$ satisfy conditions (1-2) of Theorem~\ref{thm:forward_qsp} and for all $x \in [-1, 1]$
        \begin{equation}
            |\tilde{P}(x)|^2 + (1 - x^2)|\tilde{Q}(x)|^2 \leq 1.
        \end{equation}
    The same holds if we replace real parts by imaginary parts and additionally $\tilde{P} \equiv 0$ or $\tilde{Q} \equiv 0$ can be chosen for simplicity. $\square$
\end{theorem}

While QSP is a single-qubit circuit ansatz, its mechanism can be used to great advantage in the setting of QSVT, which follows from the application of a QSP-like process within invariant two-dimensional subspaces preserved by two orthogonal projectors. While the details of this statement are left to \cite{gslw_19} and following works, QSVT should be thought of as a lifted version of QSP, where the singular values of a (near arbitrary) linear operator can be modified by a desired polynomial transformation. In this sense QSP can be thought of as the case where this linear operator is the scalar value $x$. Below we cite and simplify the presentations of the major theorems of QSVT from \cite{gslw_19}. For the reader who doesn't want to refer back to \cite{gslw_19}, we provide a short intuitive justification for the lifting argument in Remark~\ref{remark:intuitive_qsvt}.

\begin{definition}[QSVT protocol \cite{gslw_19}] \label{def:qsvt_protocol}
    Let $\Pi, \tilde{\Pi}$ be orthogonal projectors and $U$ a unitary, each acting on some $\mathcal{H}$, a finite dimensional Hilbert space. Let $n$ a positive integer be even. A QSVT protocol is the interleaved product,
        \begin{equation}
            U_{\phi} \equiv \prod_{j = 1}^{n/2} \left[e^{i\phi_{2j - 1}(2\Pi - I)}U^{\dag} e^{i\phi_{2j}(2\tilde{\Pi} - I)}U\right].
        \end{equation}
    If $n$ is odd then the protocol is only slightly modified:
        \begin{equation}
            U_{\phi} \equiv e^{i\phi_1(2\tilde{\Pi} - I)}U
            \prod_{j = 1}^{(n-1)/2} \left[e^{i\phi_{2j}(2\Pi - I)}U^{\dag} e^{i\phi_{2j+1}(2\tilde{\Pi} - I)}U\right],
        \end{equation}
    where the difference in form guarantees relation to QSP in the preserved Jordan subspaces. $\square$
\end{definition}

\begin{theorem}[QSVT and qubitization] \label{thm:qsvt_qubitization}
   Theorem 17 in \cite{gslw_19}. Let $U_{\Phi}$ a QSVT sequence with orthogonal projectors $\Pi, \tilde{\Pi}$ such that $A = \tilde{\Pi} U \Pi$ is a linear operator. Then the following relation holds:
        \begin{equation}
            P^{(SV)}(\tilde{\Pi}U\Pi) 
            =
            \begin{cases}
                \;\Pi U_{\Phi}\Pi & \text{$n$ is even,}\\
                \;\tilde{\Pi} U_{\Phi}\Pi & \text{$n$ is odd.}
            \end{cases}
        \end{equation}
    Here $P^{(SV)}(\tilde{\Pi}U\Pi) = P^{(SV)}(A)$ is the following transformation:
        \begin{equation}
            \sum_{k = 1}^{d_{\rm min}}\xi_k |\tilde{\psi}_k\rangle\langle \psi_{k}| \;\mapsto\; \sum_{k = 1}^{d_{\rm min}}P(\xi_k) |\tilde{\psi}_k\rangle\langle \psi_{k}|,
        \end{equation}
    where $P$ is a polynomial function as in standard QSP according to $\Phi$, and the $\xi_k$ are the singular values of $A$, whose left and right singular vectors are $|\tilde{\psi}\rangle, |\psi\rangle$ respectively, up to the minimum dimension of the left and right singular vector spaces. $\square$
\end{theorem}

\begin{remark}[An intuitive understanding of QSVT] \label{remark:intuitive_qsvt}
    As hinted in the previous Theorem~\ref{thm:qsvt_qubitization}, the key insight to make QSVT possible is the key observation in \cite{jordan_75} that products of two reflections or rotations preserve one and two dimensional subspaces. Consequently the action of the circuit in Def.~\ref{def:qsvt_protocol} is expressible in the subspaces mapping the right singular vectors to the left singular vectors. Careful work in \cite{gslw_19} shows that the action of $U$ with respect to this basis acts as
        \begin{equation} \label{eq:qsvt_u_decomp}
            U = \cdots \oplus\; \bigoplus_{\xi_{j} \neq 0, 1}
            \begin{bmatrix}
                \xi_j & \sqrt{1 - \xi_j^2}\;\\
                \sqrt{1 - \xi_j^2} & -\xi_j
            \end{bmatrix}^{\mathcal{H}_j}_{\tilde{\mathcal{H}}_j}
            \oplus \cdots,
        \end{equation}
    where the block's superscript $\mathcal{H}_{j}$ and subscript $\tilde{\mathcal{H}}_j$ indicate that it maps from the space spanned by the $|\psi_j\rangle$ to that spanned by the $|\tilde{\psi}_j\rangle$. Further, the two projection-dependent rotation operators, shown in \cite{gslw_19} to be easily constructable, have the block-diagonal form
        \begin{align}
            e^{i\phi(2\Pi - I)} &= \cdots \oplus\; \bigoplus_{\xi_{j} \neq 0, 1}
            \begin{bmatrix}
                e^{i\phi} & 0\;\\
                0 & e^{-i\phi}
            \end{bmatrix}^{\mathcal{H}_j}_{\mathcal{H}_j}
            \oplus \cdots,\label{eq:phase_decomp_1}\\
            e^{i\phi(2\tilde{\Pi} - I)} &= \cdots \oplus\; \bigoplus_{\xi_{j} \neq 0, 1}
            \begin{bmatrix}
                e^{i\phi} & 0\;\\
                0 & e^{-i\phi}
            \end{bmatrix}^{\tilde{\mathcal{H}}_j}_{\tilde{\mathcal{H}}_j}
            \oplus \cdots,\label{eq:phase_decomp_2}
        \end{align}
    which together with the action of $U$ allow us to recognize interleaved products of these operators as performing effectively (up to substitutions of rotations for reflections) QSP in each of the singular vector subspaces defined by these projectors. For an expedited version of this argument, based on the more concrete, albeit less general cosine-sine decomposition, we also refer readers to the excellent review in \cite{cs_qsvt_tang_tian}. $\square$
\end{remark}

\subsection{Condensed notation for QSP and QSVT} \label{sec:condensed_notation_qsp_qsvt}

Throughout the discussion of flat nesting (Def.~\ref{def:qsvt_flat_nesting}) and deep nesting (Def.~\ref{def:qsvt_deep_nesting}) of QSVT, we make use of an abridged notation for QSVT protocols, specifically devised for the aims of this paper. The intent was to more clearly manipulate the basic elements of these protocols, which are defined in terms of their phase angles $\Phi$, their pair of orthogonal projectors $\tilde{\Pi}, \Pi$, and their oracle unitary $U$. When we reference a tuple of the following form:
    \begin{equation}
        (\Phi, \tilde{\Pi}, \Pi, U),
    \end{equation}
we will always mean the QSVT protocol which has phases $\Phi$, left and right projectors $\tilde{\Pi}, \Pi$ respectively, and has oracle unitary $U$, following the original notation of \cite{gslw_19}. When one of the elements of this tuple is replaced by an asterisk $\ast$, we take the tuple to represent the unapplied function which takes as argument the missing element and outputs a QSVT circuit having performed the relevant replacement. In flat and deep nesting of QSVT protocols, we manipulate such unapplied functions using a higher-order function (nesting), which itself returns an unapplied function. Additionally, products within slots of this tuple represent standard matrix multiplications, while composition symbols, specifically $(\Phi_1 \circ \Phi_0)$ and $(\Phi_1 \land \Phi_0)$, represent simple algorithms taking two lists of QSVT phases and returning a single (longer) list of QSVT phases obeying the relevant composition algorithm (in this case flat and deep nesting respectively).

While this work makes many references to the standard composition symbol, e.g., $(\Phi_1 \circ \Phi_0)$ and $(G \circ F)$, we have tried to provide ample context, based primarily on the objects considered in the composition, for exactly what is being meant by the symbol.

\subsection{Imposing symmetries on QSP phase lists} \label{sec:constrained_qsp_phases}

In this section we briefly cover basic properties of QSP protocols which have special symmetries imposed on their phase lists. We phrase the action of these symmetries in terms of a group action induced on tuples of polynomials (identical to those embedded by the relevant QSP circuits). Interest in such `symmetrized QSP protocols' has grown recently \cite{sym_qsp_21}, and may have wider utility in the theory of QSP/QSVT than either numerical stability or semantic embedding. It is an open question whether there exist further (i.e., beyond those given here) easily expressible manipulations of QSP phases which preserve, independently, $|P|^2$ and $|Q|^2$ as defined in the main lemma of this section. For instance, does there exist a simple transformation on the QSP phases of a protocol which embeds, not $P$, but $P$ with one of its roots taken to its complex conjugate (along with modifications to other roots to preserve define parity).

\begin{definition}[Group action]
    Given a group $G$ and a set $X$, a (left) group action $\zeta$ of $G$ on $X$ is a function $\zeta : G \times X \mapsto X$ satisfying the following two properties: (1) $\forall x \in X, \zeta(e, x) = x$, and (2) $\forall g, h \in G, \forall x \in X, \zeta(g, \zeta(h, x)) = \zeta(gh, x)$. One can also think of this as a group homomorphism from $G$ to the group of bijections from $X$ to itself. $\square$
\end{definition}

\begin{lemma}[Group actions on QSP phases] \label{lemma:protocol_symmetries}
    Given a QSP protocol $U_{\Phi}$ defined equivalently by phases $\Phi = \{\phi_0, \phi_1, \cdots, \phi_n\}$ and two complex polynomials $P, Q$ satisfying $|P|^2 + (1 - x^2)|Q|^2 = 1$, the two discrete transformations of $\Phi$
        \begin{align}
            R: \Phi &\longmapsto \{\phi_n, \phi_{n - 1}, \cdots, \phi_1\}\; \text{ (Reverse)},\\
            N: \Phi &\longmapsto -\Phi\;\text{ (Negate)},
        \end{align}
    act on $(P, Q)$ according to the following diagram:
    \begin{equation}
        \begin{tikzcd}
            (P, Q) \ar[r,"R"] \ar[d,"N"] & (P, -Q^*) \ar[d,"N"]\\
            (P^*, -Q^*) \ar[r, "R"] & (P^*, Q).
        \end{tikzcd}
    \end{equation}
    In other words, we have identified a group action for $(\mathbb{Z}_2\times \mathbb{Z}_2)$ on pairs of polynomials.
    
    One can also consider the following two additional discrete transformations of $\Phi$:
        \begin{align}
            A: \Phi &\longmapsto (\phi_1 + \pi/2, \phi_2, \cdots, \phi_{n} -\pi/2)\; \text{ (Antisymmetric)},\\
            S: \Phi &\longmapsto (\phi_1 + \pi/2, \phi_2, \cdots, \phi_{n} +\pi/2)\; \text{ (Symmetric)},
        \end{align}
    where we have added (or subtracted) said constant to each phase in the list. Simple computation reveals that this generates the following transformation on embedded polynomials
        \begin{equation}
            \begin{tikzcd}
                (P, Q) \ar[r,"A"] \ar[d,"S"] & (P, -Q) \ar[d,"S"]\\
                (-P, Q) \ar[r, "A"] & (-P, -Q).
            \end{tikzcd}
        \end{equation}
    It is not so difficult to see that $N, R, A, S$ form a group action of $(\mathbb{Z}_2)^{\times 4}$ on the set $\{\pm P, \pm P^{*}\}\times \{\pm Q, \pm Q^{*}\}$, and that this action is transitive on the set (by their identical size). We will make use of these symmetries when discussing specific QSP protocols in the following definitions. Note that all group elements preserve the required $|P|^2 + (1 - x^2)|Q|^2 = 1$, and in fact preserve $|P|^2$ and $|Q|^2$ independently. $\square$
\end{lemma}

\section{Proofs for semantic embedding for QSP and QSVT} \label{sec:embedded_qsp_qsvt}

In this appendix we provide proofs of the theorems given in Sec.~\ref{sec:semantic_embedding_qsp} and Sec.~\ref{sec:semantic_embedding_qsvt}. We again assume basic familiar with the methods of proof for standard QSP, which are covered carefully in \cite{gslw_19}, and more loosely in Appendix~\ref{sec:qsp_qsvt_theory}.

\subsection{Semantic embedding for QSP}

\begin{remark}[Proof of Theorem~\ref{thm:existence_uniqueness_asym_qsp}]
    We break this proof into its two major components, the existence of a set of antisymmetric phase factors provided a unitary satisfies the stated conditions, and the uniqueness of these phase factors up to a natural restriction of their domain.
    
    (Existence.) Take $U_{\Phi}$ with $P, Q$ satisfying the conditions of Theorem~\ref{thm:existence_uniqueness_asym_qsp}. To show the existence of a $\Phi$ satisfying the desired antisymmetric properties we consider attempting to lower the degree of the polynomials embedded in $U_{\Phi}$ according to the application of the inverse of an antisymmetric QSP iterate. In other words the existence of a $\phi$ such that
        \begin{equation}
            W^\dagger e^{-i\phi \sigma_z} U_{\Phi} e^{i\phi \sigma_z} W^\dagger  = 
            W^{\dagger} e^{-i\phi\sigma_z}
            \begin{bmatrix}
                P & Q\sqrt{1 - x^2}\\
                Q^* \sqrt{1 - x^2} & P
            \end{bmatrix}
            e^{i\phi\sigma_z} W^{\dagger},
        \end{equation}
    encodes some $P^\prime, Q^\prime$ also satisfying the hypotheses of the theorem statement, in addition to having strictly lower degree. The application of this iterate transforms $P, Q$ into the new $P^\prime, Q^\prime$ with the forms
        \begin{align}
            P^\prime &= x(1 - x^2)(e^{-2i\phi}Q + e^{2i\phi}Q^*) + P(2x^2 - 1)\label{eq:phi_conditions_1}\\
            Q^\prime &= -2P x - e^{2i\phi} Q^* (1 - x^2) + x^2 e^{-2i\phi} Q.\label{eq:phi_conditions_2}
        \end{align}
    This shows that evidently $P^\prime$ is real, as required, as well as that the parity constraints are respected. The determinantal constraint is also assured by the unitarity of the overall transformation. To show the non-trivial result, namely that there exists a choice of $\phi$, and specifically the $\phi$ with
        \begin{equation} \label{eq:special_phi}
            e^{2i\phi} = P_{d}/Q_{d - 1}^*,
        \end{equation}
    such that the overall degree of the embedded polynomials is lowered, we need to show that this ratio both has unit modulus, as well as satisfies that its insertion into Eqs.~\ref{eq:phi_conditions_1} and \ref{eq:phi_conditions_2} leads to both $P^\prime_{d + 2}$ and $P_{d}$ being zero (as well as the corresponding leading coefficients of $Q^\prime$). The first part of this is easy, as $P_d = |Q_{d-1}|$ again by the unitarity of the transformation. The second set of conditions breaks down into individual constraints on the leading coefficients, which we give here
        \begin{align}
            P^\prime_{d + 2} = 2P_{d} - \Re[e^{-2i\phi} Q_{d - 1}] &= 0\\
            P^\prime_{d} = -P_d + 2P_{d - 2} + \Re[e^{-2i\phi}Q_{d - 1}] - \Re[e^{-2i\phi}Q_{d - 3}] &= 0,
        \end{align}
    The first equation is evidently satisfied by our choice of $\phi$ in Eq.~\ref{eq:special_phi}, and so what remains is to show that the second is also satisfied through properties of the determinant of a unitary embedding polynomials of definite degree and parity. We can use the first equation to simplify the second, recovering
        \begin{equation} \label{eq:sub_leading_condition}
            P^\prime_{d} = P_d + 2P_{d -2} - \Re[e^{-2i\phi}Q_{d - 3}] - \big(2P_d - \Re[e^{-2i\phi}Q_{d - 1}]\big) = 0,
        \end{equation}
    which is shown to be true for unitary matrices satisfying the parity and degree constraints given in the theorem statement by the following short Lemma~\ref{lemma:sub_leading_unitary_conditions}.
        
    \begin{lemma}[Conditions on antisymmetric QSP polynomials] \label{lemma:sub_leading_unitary_conditions}
        For a unitary matrix satisfying the conditions of Theorem~\ref{thm:asym_embedded_qsp_equivalence}, the coefficients of $P, Q$ satisfy not only the leading condition $P_{d} = |Q_{d - 1}|$ but the sub-leading condition
            \begin{equation}
                2P_{d - 2} + |Q_{d - 1}| - \Re[e^{-2i\phi}Q_{d - 3}].
            \end{equation}
        for some real choice of $\phi$. Proof is by observing the determinantal equation
            \begin{equation}
                P(x)^2 + (1 - x^2)e^{2i\phi} Q(x) e^{-2i\phi}Q(x) = 1,
            \end{equation}
        and setting its $x^{2d}$ and $x^{2d - 2}$ coefficients to zero. This leaves the expression
            \begin{equation}
                2 P_{d} P_{d - 2} + |Q_{d - 1}|^2 - |Q_{d - 1}|\Re[e^{-2i\phi}Q_{d - 3}],
            \end{equation}
        which we can simplify by removing a factor of $|Q_{d - 1}|$ using the guaranteed unitarity condition into the form given, i.e., that $|Q_{d-1}| = Q_{d - 1}e^{2i\phi}$. Converting to the form in Eq.~\ref{eq:sub_leading_condition} just involves replacing $|Q_{d-1}|$ with $P_{d}$, again by unitarity. The condition given on the leading coefficient of $P$ in the case that $d$ is even is proven in precisely the same way as Lemma 10 in \cite{sym_qsp_21}, save with $\phi_k = -\phi_{n - k}$ used instead of their symmetric condition, which does not modify the sign of the cosine terms involved there. $\square$
    \end{lemma}
    
    (Uniqueness.) The uniqueness result will follow from both existence (given above), and an inductive argument on the antisymmetric phases defining the desired QSP protocol. Note that in the base case, when $P_{d}$ for $d \geq 2$ is equal to zero, that $Q_{d - 1}$ must also be identically zero, and thus the unitary $U_\Phi$ is the identity, which is achieved as stated above by the trivial antisymmetric phase list $\Phi = \{0\}$.
    
    Assume toward contradiction that there exists two distinct $\phi$ such that the following equality holds
        \begin{equation}
            W^\dagger e^{-i\phi \sigma_z} U_{\Phi} e^{i\phi \sigma_z} W^\dagger  = 
            W^{\dagger} e^{-i\phi\sigma_z}
            \begin{bmatrix}
                P & Q\sqrt{1 - x^2}\\
                Q^* \sqrt{1 - x^2} & P
            \end{bmatrix}
            e^{i\phi\sigma_z} W^{\dagger}
        \end{equation}
    where we denote the resulting unitary by the following
        \begin{equation}
            U_{\Phi^\prime} = 
            \begin{bmatrix}
                P^\prime & Q^\prime\sqrt{1 - x^2}\\
                (Q^\prime)^* \sqrt{1 - x^2} & P^\prime
            \end{bmatrix},
        \end{equation}
    where $P, Q$ satisfy the conditions given in the statement of Theorem~\ref{thm:existence_uniqueness_asym_qsp} for degree, parity, and realness (i.e., the inductive hypothesis). By existence of a valid set of antisymmetric phase factors, both $\phi, \phi^\prime$ must result in $P^\prime, Q^\prime$ having degree strictly less than that of $P, Q$ (excluding the inductive base case). This condition is the same as the following condition on the leading coefficients of $P, Q$
        \begin{equation}
            P^\prime_{d + 2} = 2P_{d} - e^{-2i\phi} Q_{d - 1} - e^{2i\phi} Q_{d - 1}^* = 0,
        \end{equation}
    which, along with the additional known determinantal constraint that $P_{d} = |Q_{d - 1}|$ and $P_d = P_d^*$, leads to the following expression for $\phi$,
        \begin{equation}
            e^{2i\phi} = P_{d}/Q_{d-1}^*,
        \end{equation}
    which, as in the analogous proof of \cite{sym_qsp_21} for their symmetric protocol, has only one real solution $\phi$ within the domain $D_{d}$. This process can be repeated, and shows uniqueness of $\Phi$ by induction, following from the existence result proved earlier.
\end{remark}

\begin{remark}[Proof of Theorem~\ref{thm:asym_embedded_qsp_equivalence}]
    The proof of this statement, that antisymmetric QSP protocols are in bijection with embeddable QSP protocols, is relatively straightforward. We appeal to the standard representation in terms of generators of SU(2) of rotations on the Bloch sphere. Namely, for the map from antisymmetric protocols to embeddable ones take
        \begin{align}
            f(x) &= [1 - P(x)^2]^{-1/2}\Re(Q(x))\sqrt{1 - x^2},\\
            g(x) &= [1 - P(x)^2]^{-1/2}\Im(Q(x))\sqrt{1 - x^2}.
        \end{align}
    For the reverse mapping we perform a similar operation,
        \begin{align}
            P(x) &= [1 - x^2]^{-1/2}\sqrt{f(x)^2 + g(x)^2},\\
            Q(x) &= [1 - x^2]^{-1/2}[f(x) + i g(x)].
        \end{align}
    Note that parity constraints, norm constraints, and $f(\pm1) = g(\pm 1) = 0$ are also preserved by properties of $P, Q, f, g$ as given. This choice is unambiguous by the uniqueness of $P, Q$ given $\Phi$ (up to domain restriction) as per Def.~\ref{def:asym_qsp_protocol} and proven in Theorem~\ref{thm:existence_uniqueness_asym_qsp}, and thus the protocols are in bijection.
\end{remark}

\begin{remark}[Proof of Theorem~\ref{thm:partial_asym_qsp}]
   The proof of this statement is related to similar results on matrix completions in \cite{gslw_19, haah_2019, ylc_14, lyc_16_equiangular_gates, lc_17_simultation, lc_19_qubitization} and so on. The modification we provide hinges on the assumed realness of $P$ (following in turn from the antisymmetry of the QSP phases).
    
    Concretely, we require that the partially specified unitary transform
        \begin{equation}
            U = 
            \begin{bmatrix}
                P & \cdot \\
                \cdot & P
            \end{bmatrix}
        \end{equation}
    can be \emph{completed} (i.e., the dots filled in) such that these additional elements satisfy properties (1-5) in the theorem statement as well as that the overall matrix remains unitary. If this is the case then Theorem~\ref{thm:existence_uniqueness_asym_qsp} can be invoked, and basic classical algorithms can be used to efficiently determine the required antisymmetric QSP phases from the coefficients of $P$.
    
    In this case, the only new condition is (4), as condition (5) will follow directly from invoking Lemma 10 of \cite{sym_qsp_21} as well as the uniqueness property of Theorem~\ref{thm:existence_uniqueness_asym_qsp}. However, condition (4) has evident meaning. Note that the definite parity of $P$ as well as its realness for $x \in [-1,1]$ requires that its roots come in pairs $(\alpha, -\alpha)$ and $(\alpha, \alpha^*)$, or equivalently that they are closed under both negation and complex conjugation. In order that $Q$ has the properties required by Theorem 5 in \cite{gslw_19}, we need that $1 - P(x)^2 = |Q(x)|^2$ induces $Q(x)$ to be of definite parity. As this is a product of a polynomial with its complex conjugate, it helps to transform to the Laurent picture $x \mapsto (z + z^{-1})$. Evidently $1 - P([z + z^{-1}]/2)^2$ is a polynomial which is real and non-negative for all $z \in \mathbb{T}$ (the complex numbers of modulus one). 
    
    Note that condition (4) is strictly stronger than what is implied by the Fejér-Riesz theorem as quoted in \cite{rossi_m_qsp_22}; this theorem only guarantees the existence of a $Q(x)$ such that $1 - P(x)^2 = |Q(x)|^2$. Namely, while $1 - P(x)^2$ is positive and thus can always be factored into some $|Q(x)|^2$, the parity of $Q(x)$ is not definite unless the roots of $Q(x)$ after transforming to the Laurent picture, i.e.,
        \begin{equation} \label{eq:laurent_factorization}
            |Q(x)|^2 = \alpha \prod_{r_i} (z - r_i)(z^{-1} - r_i),
        \end{equation}
    are such that the multiset of $r_i$ is closed under negation. This property is necessary and sufficient for a polynomial to have definite parity in both the Laurent and $x$ picture, as sending $z \mapsto -z$ or $x \mapsto -x$ necessarily adds only a sign to Eq.~\ref{eq:laurent_factorization} under the assumption of closure.
\end{remark}

\begin{remark}[Proof of Theorem~\ref{thm:asym_qsp_composition}]
    The proof of this theorem follows mostly directly from the properties of twisted oracles discussed in Lemma~\ref{lemma:twisted_qsp}. That is, the realness of $P$ for antisymmetric QSP protocols corresponds to the fact that, for any individual $x$, the oracle provides a rotation about some axis on the equator of the Bloch sphere. This is, as noted, equivalent to the definition of embeddable QSP protocols in Def.~\ref{def:embeddable_protocols}. More importantly, as conjugation of the signal oracle by $\sigma_z$-rotations does not modify the QSP phases save on the extreme ends of the QSP phase list, the functional transform applied by the outer protocol $\Phi_1$ is necessarily applied to the result of the inner protocol $\Phi_0$. In other words, the inner protocol provides an oracle which looks like
        \begin{equation}
            U_{\Phi_0} = e^{iG(x)\sigma_z}W(P_0(x))e^{-iG(x)\sigma_z},
        \end{equation}
    where $W(x) = e^{i\arccos{x}\sigma_x}$ as before, and the rotation $G(x)$ can be written explicitly in the following $x$-dependent form (among many equivalents)
        \begin{equation}
            G(x) = \arccos{[\Re({Q(x)})/|Q(x)|]}.
        \end{equation}
    As discussed in Lemma~\ref{lemma:twisted_qsp}, this conjugation does not affect the ability for ab outer QSP protocol to apply a desired functional transform to the on-diagonal elements of $U_{\Phi_0}$ in the computational basis for any $x$, and thus the composite protocol $U_{\Phi_1 \circ \Phi_0} = (\Phi_1 \circ \Phi_0)(x)$ enacts $(P_1 \circ P_0)$ as intended.
\end{remark}

\subsection{Semantic embedding for QSVT}

\begin{remark}[Proof of Theorem~\ref{thm:qsvt_flat_embedding}] \label{proof:qsvt_flat_embedding}
    Proof of the properties of flatly embedded QSVT is straightforward, recalling Eq.~\ref{eq:qsvt_u_decomp}, which states that the action of the QSP unitary is, in each of the preserved two-dimensional subspaces from Jordan's lemma,
        \begin{equation}
            U = \cdots \oplus\; \bigoplus_{\xi_{j} \neq 0, 1}
            \begin{bmatrix}
                \xi_j & \sqrt{1 - \xi_j^2}\;\\
                \sqrt{1 - \xi_j^2} & -\xi_j
            \end{bmatrix}^{\mathcal{H}_j}_{\tilde{\mathcal{H}}_j}
            \oplus \cdots.
        \end{equation}
    Note that if this unitary is given as the unitary oracle for another QSP protocol using the same projectors and computing registers, that the action is identical to that of QSP in each qubit-like subspace assured by Jordan's lemma. Consequently the conditions under which QSP can compose (Theorem~\ref{thm:asym_qsp_composition}) are precisely those under which flatly embedded QSVT is possible, and that the QSVT protocol remains flatly embeddable is also a consequence of this argument. As shown in the following result, however, the presence of additional freedoms in defining a QSVT protocol permits further semantic embedding techniques which have no direct equivalence to composition of QSP within each invariant qubit-like subspace.
\end{remark}

\begin{remark}[Proof of Theorem~\ref{thm:qsvt_deep_embedding}] \label{proof:qsvt_deep_embedding}
    For deep embedding in QSVT, the inner protocol functions identically to that of standard QSVT, as their is no restriction on the QSVT phases used. In this case, we take that this protocol block-encodes the transformed scalar value
        \begin{equation}
            \tilde{\Pi}_0 U_0 \Pi_0 = \sqrt{F(|A_0|)},
        \end{equation}
    where $\tilde{\Pi}_0\Pi = |A_0|$ is the analogue of the overlap between the uniform superposition and the marked subspace in Grover search. Deep nesting in QSVT refers to the inner protocol being used to conjugate one of the projectors (the marking projector $\tilde{\Pi}_1$) of the outer protocol. In this case the action of this conjugation is to produce a projector which is both in the image of $\tilde{\Pi}_0$ (up to the factor $\sqrt{F(|A_0|)}$) as well as overlapping with the defining projector $\tilde{\Pi}_1$ of the outer protocol (up to the factor $\sqrt{F(|A_1|)}$). Consequently the overall action of the circuit is to block-encode the product of these two functional transforms in its upper-left block, as given in the theorem statement. When $F, G$ each approximate step functions at the standard Grover threshold:
        \begin{equation}
            F(x) \approx_{\varepsilon} \Theta(x - 1/\sqrt{|A_0|}),
            \quad\quad
            G(x) \approx_{\varepsilon} \Theta(x - 1/\sqrt{|A_1|}),
        \end{equation}
    then the action of the overall circuit becomes the logical AND of the set membership oracle implicitly defined by projectors $\tilde{\Pi}_0, \tilde{\Pi}_1$. That this product can also be made to $\varepsilon$-approximate this AND function follows from theorems in \cite{gslw_19} on the product of block-encoded operators.
\end{remark}

\section{Basics of category theory} \label{sec:cat_theory}

We present a minimal series of definitions of categories, functors (morphisms between categories), and natural transformations (morphisms between functors). We use these definitions to concretely indicate what is meant by semantically embedded QSP. Nearly all quoted results follow the well-known introductory textbook \cite{cat_theory_78} on category theory.

\subsection{Categories}

Category theory is founded on the observation that various mathematical properties are unified and simplified through presentation in terms of diagrams of arrows. Here arrows $f: X \rightarrow Y$ represent functions from sets $X$ to sets $Y$. Such diagrams are said to be \emph{commutative} when all valid (respecting the direction of arrows) paths with the same start and end points are equal. Such diagrams can be used to depict countless common mathematical structures in a way agnostic to their instantiation; the utility of category theory is often attributed to the ability of diagrams to vividly represent the action of their arrows.

We hold off on formally defining categories (which is somewhat technical, and involves first defining metacategories, their axioms, and then restricting such axioms to within set theory). For our purposes we can instead define \emph{graphs} as a set of objects $O$ and arrows $A$, as well as functions from $A$ to $O$
    \begin{equation}
        \begin{tikzcd}
            A \arrow[r, shift right, swap, "\text{cod}"] \arrow[r, shift left, "\text{dom}"] & O.
        \end{tikzcd}
    \end{equation}
These graphs can have composable pairs of arrows, namely those arrows $g, f$ which are both in $A$ and are such that the domain of $g$ is the codomain of $f$, permitting $g \circ f$ to make sense. For categories, we thus start with graphs and add two additional functions called \emph{identity} and {composition}. 
    \begin{equation}
        \begin{tikzcd}
            O \arrow[r, "\text{id}"] & A & A\times A \arrow[r, "\circ"] & A
        \end{tikzcd}.
    \end{equation}
These functions are defined over a (directed) graph whose nodes are objects and whose arrows are arrows, such that 
    \begin{equation}
        \text{dom}(\text{id}\,a) = a = \text{cod}(\text{id}\,a), 
        \quad
        \text{dom}(g\circ f) = \text{dom}\,f,
        \quad
        \text{cod}(g\circ f) = \text{cod}\,g.
    \end{equation}
Normally we will not refer to the underlying graphs or sets of arrows and objects individually, writing $c \in C$ and $f \text{ in } C$ for objects and arrows respectively. From only this basic definition, common mathematical objects, e.g., monoids, groups, matrices, and sets, can be defined simply in terms of the commutative diagrams they satisfy.

\subsection{Functors}

A functor is a \emph{morphism of categories}. For categories $C$ and $B$, a functor, denoted $T: C \rightarrow B$ with domain $C$ and codomain $B$ consists of two related functions. There is the \emph{object function} $T$ that assigns each object $c$ of $C$ an object $Tc$ of $B$, and an \emph{arrow function} (also called $T$) that assigns arrows $f: c \rightarrow c^\prime$ of $C$ arrows $Tf: Tc \rightarrow Tc^\prime$ of $B$ such that the following hold:
    \begin{equation}
        T(1_c) = 1_{Tc} \quad\quad T(g\circ f) = Tg\circ Tf.
    \end{equation}
In other words, the functor respects the identity function in $C$, as well as valid compositions of arrows in $C$. A functor can also be defined purely in terms of arrows: $T$ is a function between arrows $f$ of $C$ to arrows $Tf$ of $B$ respecting the identity and composition properties of arrows in $C$. In casual terms, a functor translates a diagram (respecting its structure) from $C$ to $B$.

We can list a few common properties and types sometimes attributed to functors. We say a functor is \emph{forgetful} if its action is to simply forget some or all of the structure of the underlying algebraic object (for instance, the group structure assigned to the set of elements of a group). A functor which is a bijection on both arrows and objects is an \emph{isomorphism of categories}. Functors can also be composed, and this composition is associative; moreover a functor $T: C \rightarrow B$ is an isomorphism of categories if and only if there exists a functor $S: B \rightarrow C$ such that both composites $S \circ T$ and $T\circ S$ are the identity functors. Weaker than isomorphisms are \emph{full} functors, for which for every pair $c, c^\prime$ of objects and arrows $g: Tc \rightarrow Tc^\prime$ of $B$ there exists an arrow $f: c \rightarrow c^\prime$ of $C$ with $g = Tf$ (in other words, arrows in $B$ have preimages). Finally, a functor $T: C \rightarrow B$ is \emph{faithful} when every pair of objects $c, c^\prime$ of $C$ and every pair $f_1, f_2 : c \rightarrow c^\prime$ is such that $Tf_1 = Tf_2: Tc \rightarrow Tc^\prime$ implies that $f_1 = f_2$. In other words, identity of arrows in the image of the functor imply identity of arrows in the preimage.

It is worthwhile to discuss the properties of the functors defined in the assignments for the diagram in Fig.~\ref{fig:qsp_as_natural_transformation}, i.e., in Tables~(\ref{tab:qsp_categories}-\ref{tab:qsvt_deep_categories}). In the theorem below we quickly show that, under a suitable and reasonable restriction on the permitted phase lists, these functors are all both full and faithful. Before this, we give a quick definition.

\begin{definition}[Trivial and honest QSP phase lists] \label{def:honest_phases}
    A QSP phase list of length $n$ is termed \emph{honest} if the degree of the polynomial in the top-left element of the resulting unitary has degree exactly $n - 1$. Such phase lists necessarily do not contain \emph{trivial} sublists, i.e., QSP phases whose corresponding unitary is the identity. An example of such a phase list is the following:
        \begin{equation} \label{eq:trivial_concatenation}
            \Phi \cup \Phi^{RN},
        \end{equation}
    for an arbitrary QSP phase list $\Phi$, where we have joined it to its reversed and negated version, using notation from Appendix~\ref{sec:constrained_qsp_phases}. In fact, it can be shown that this operation, along with concatenation, generates all possible trivial QSP phase lists. For our purposes we will restrict to considering only honest QSP phase lists, where this property can be determined in linear time in the number of phases by looking for the presence of $\pi/2$ in the relevant phase list. $\square$
\end{definition}

\begin{theorem}[Properties of functors in assignments for Fig.~\ref{fig:qsp_as_natural_transformation}]
    Under the restriction to honest QSP phase lists (Def.~\ref{def:honest_phases}), the assignments given in Table~\ref{tab:qsp_categories} for the functors $S, T$ are both full and faithful. That is, these functors are bijective on hom-sets, or the set of arrows, e.g., $f, g, h$, in the category of nested QSP phase lists, e.g., $a, b, c$. It is worthwhile to remember that the functional transforms we consider, given by the second functor $T$, are contstrained in norm and parity, as is standard in QSP.
    \begin{proof}
        Proof follows by looking at the action of an arrow under the functor. Take $\Phi_0, \Phi_1$ honest phase lists, and consider the arrow which takes $\Phi_0$ to $\Phi_1 \circ \Phi_0$ according to Def.~\ref{def:nested_protocols}. 
        
        Considering $S$, we show its injectivity and surjectivity on hom-sets. Under the functor $S$ this corresponds to taking $U_{\Phi_0}$ to $U_{(\Phi_1 \circ \Phi_0)}$. Injectivity of $S$ follows from our restriction that $\Phi_0, \Phi_1$ are honest, as if there existed a $\Phi_1^\prime$ such that
            \begin{equation}
                U_{(\Phi_1 \circ \Phi_0)} = U_{(\Phi_1^\prime \circ \Phi_0)},
            \end{equation}
        then $U_{(\Phi_1 \circ \Phi_0)}U^{-1}_{(\Phi_1 \circ \Phi_0)}$ would be the identity, making the following concatenation trivial:
            \begin{equation}
                (\Phi_1 \circ \Phi_0) \cup (\Phi_1^\prime \circ \Phi_0)^{RN}.
            \end{equation}
        But if this phase list is trivial, and moreover $\Phi_1^\prime \neq \Phi_1$, then in the process of decomposing this trivial phase list into a series of concatenations (\ref{eq:trivial_concatenation}), as must be possible as each half of the above expression is honest, there must be two non-identical sublists which evaluate to the same unitary, contradicting the assumed honesty of the phases $\Phi_0, \Phi_1$. For surjectivity of $S$, the proof is simpler, as QSP says that all validly specified QSP unitaries have a corresponding QSP phase list.

        For the functor $T$, the relevant action takes a function $f$ to the composition $g\circ f$; that this functor is injective on arrows follows from the observation that honest phase lists which encode the same functional transform must be the same, as otherwise their concatenation $\Phi \cup (\Phi^{\prime})^{RN}$ would be both trivial and non-empty, contradicting the honesty of at least one of the protocols. Surjectivity of $T$ follows from the computable map between functional transforms and QSP protocols.
    \end{proof}
\end{theorem}

\subsection{Natural transformations}

Given two functors $S, T: C \rightarrow B$, a natural transformation $\tau: S \rightarrow T$ is a function that assigns objects $c$ of $C$ arrows $\tau c : Sc \rightarrow Tc$ of $B$ such that every arrow $f: c\rightarrow c^\prime$ in $C$ yields a diagram that is commutative (see Fig.~\ref{fig:natural_transformation_def}).

\begin{figure}[htpb]
    \centering
    \begin{tikzcd}
        c \arrow[d, "f"] & Sc \arrow[d, "Sf"] \arrow[r, "\tau c"] & Tc \arrow[d, "Tf"]\\
        c^\prime & Sc^\prime \arrow[r, "\tau c^\prime"] & Tc^\prime
    \end{tikzcd}
    \caption{The standard depiction of a natural transformation, given two functors $S, T$. The components of the natural transformation $\tau$ respect all arrows in the original category.}
    \label{fig:natural_transformation_def}
\end{figure}

When this holds, it is said that $\tau c$ is natural in $c$. A natural transformation $\tau$ can be viewed as a set of arrows mapping the `picture' $S$ to the `picture' $T$ (taking that $S$ provides a picture in $B$ of the objects and arrows of $C$). These pictures can in general be made more complex, as in Fig.~\ref{fig:qsp_as_natural_transformation}. In all cases (e.g., Fig.~\ref{fig:qsp_as_natural_transformation}), the $\tau a, \tau b, \tau c$ are called the components of the natural transformation $\tau$. One can also define natural transformations as \emph{morphisms of functors}.


\end{document}